\documentclass[preprints,review,accept,moreauthors,pdftex]{mdpi}

\usepackage{graphicx}
\usepackage{here}
\usepackage{stackrel}
\usepackage{amssymb,latexsym}
\usepackage{braket}

\usepackage{subfigure}
\usepackage{tikz}

\newcommand{\B}{{\rm{B}}}

\newcommand{\Tr}{{\textrm{Tr}}}

\newcommand{\MeV}{\,{\rm MeV}}
\newcommand{\GeV}{\,{\rm GeV}}

\newcommand{\QCD}{{\textrm{\scriptsize QCD}}}

\newcommand{\TF}{{\textrm{\scriptsize TF}}}
\firstpage{1} 
\pubvolume{xx}
\issuenum{1}
\articlenumber{5}
\pubyear{2020}
\copyrightyear{2020}
\history{Received: 10 August 2020; Accepted: 30 August 2020; Published: date}
\updates{yes} % If there is an update available, un-comment this line

\Title{Fractal Structures of Yang--Mills Fields and Non-Extensive Statistics: Applications to \mbox{High Energy Physics}}%Attention AE/ME. The following layout issues have not been checked by the English Editing Department and must be carefully verified by the AE/Layout Department: All callout issues, bold usage of callouts, and references to callouts in the text. Correct callout usage in figures. Figure and Table layout issues. Footnote formatting and Glossaries have not been checked. En dash usage for negative values, en dash usage to indicate relationships, en dash usage to indicate bonds (especially in chemistry). The English Editing Department is not responsible for correct italic usage for genes, proteins and technical terminology. This responsibility belongs to the authors. The following are also not checked: spacing between numbers and units of measurement, ratios, en dashes for ranges, date and time formats, punctuation in equation lines, and less than/more than spacing (< >). Finally, capitalization and layout of titles/headings must be properly checked as well as ensuring 'Eq.' and 'Fig.' are properly spelled out, as these are layout issues.

\Author{Airton Deppman $^{1,\dagger}$\orcidA{}, Eugenio Meg\'{\i}as $^{2,}$*$^{,\dagger}$\orcidB{} and D\'ebora P. Menezes $^{3,\dagger}$\orcidC{}}

\AuthorNames{Airton Deppman, Eugenio Meg\'{\i}as and D\'ebora P. Menezes}

\address{%
$^{1}$ \quad Instituto de F\'{\i}sica, Rua do Mat\~ao 1371-Butant\~a, S\~ao Paulo-SP, CEP 05580-090, Brazil; deppman@if.usp.br\\
$^{2}$ \quad Departamento de F\'{\i}sica At\'omica, Molecular y Nuclear and Instituto Carlos I de F\'{\i}sica Te\'orica y Computacional, Universidad de Granada, Avenida de Fuente Nueva s/n, 18071 Granada, Spain\\
$^{3}$ \quad Departamento de F\'{\i}sica, CFM-Universidade Federal de Santa Catarina, Florian\'opolis, SC-CP. 476-CEP 88.040-900, Brazil; debora.p.m@ufsc.br}

\corres{Correspondence:
emegias@ugr.es; Tel.: +34-958-241000}

\abstract{In this work, we provide an overview of the recent
investigations on the non-extensive Tsallis statistics and its
applications to high energy physics and astrophysics, including
physics at the Large Hadron Collider (LHC), hadron physics, and neutron stars. We review some recent investigations on the power-law distributions arising in
high energy physics experiments focusing on a thermodynamic description of the system formed, which~could explain the power-law behavior. The
possible connections with a fractal structure of hadrons is also
discussed. The main objective of the present work is to delineate
the state-of-the-art of those studies and~show some open issues
that deserve more careful investigation. We propose several
possibilities to test the theory through analyses of experimental
data. } 


\begin{document}
%%%%%%%%%%%%%%%%%%%%%%%%%%%%%%%%%%%%%%%%%%

%%%%%%%%%%%%%%%%%%%%%%%%%%%%%%%%%%%%%%%%%%

\section{Introduction}
\label{sec:introduction}

An interesting description of hadronic systems in the hot and in the dense regimes, known as quark-gluon plasma (QGP), has been developed in recent years. Although motivated by the large amount of information that emerged from high energy physics (HEP) experiments, the consequences of those advances are far-reaching, since they may be present in any Yang--Mills field (YMF) theory. Here, we will give a short review of those developments, discuss the experimental evidence of the new theoretical approach, and present some applications to HEP, hadron physics, and astrophysics.

Before entering in the main subject of this work, let us summarize the three fundamental theories that form the foundations for the developments that will be discussed below. These three theories are: the YMF; the fractal geometry; and the non-extensive statistics proposed by Constantino Tsallis. 

The Yang--Mills field theory is a prototype theory for describing most
of the physical phenomena. It~was proposed by Yang and Mills in the
early 1950s~\cite{Yang:1954ek} and was incorporated in the description
of interacting fields that is known as Standard Model: it was
incorporated into the electro-weak theory in the 1960s and in quantum
chromodynamics (QCD) in the 1970s. Basically, it describes the
propagation of undulatory fields in space and time, as~well as the
interaction between those fields. One of the fundamental properties of
physics laws is the renormalization group invariance, and~all the four
known fundamental interactions show this feature as an essential
aspect of its structure. At least three of them can be described by
YMF,
%what gives to those interactions another special feature, the complexity of the fields, 
but for the gravitational interaction, a YMF version \mbox{has also been proposed}. 

The simplest non-Abelian gauge field theory whose Lagrangian density includes bosons and fermions is given by

\begin{equation}
{\cal L}=-\frac{1}{4} F^{a}_{\mu \nu} F^{a \, \mu \nu}+ i \bar{\psi_i} \gamma_{\mu} D^{\mu}_{ij} \psi_j \,, \label{YMLagrangean}
\end{equation}
where~$F^{a}_{\mu \nu}=\partial_{\mu} A^{a}_{\nu} - \partial_{\nu} A^{a}_{\mu}+g f^{abc}A^{b}_{\mu}A^{c}_{\nu}$ is the field strength of the gauge field, and~$D^{\mu}_{ij}= \partial_{\mu}\delta_{ij}-igA^{a \, \mu} T^{a}_{ij}$ is the covariant derivative, with $f^{abc}$ being the structure constants of the group, and~$T^{a}$ the matrices of the group generators. In these expressions, $\partial_\mu = \partial/\partial x^\mu$ with $x^\mu$ $(\mu=0,1,2,3)$ the spacetime coordinates, $a = 1, \cdots, N_c^2-1,$ 
are the color group indices with $N_c$ the number of colors, $g$ is the coupling constant, $\gamma_\mu$ $(\mu=0,1,2,3)$ are the Dirac matrices, and $i,j = 1, \cdots, N_f$, are flavor indices with $N_f$ the number of quark flavors. $\psi$ and $A$ represent, respectively, the fermion and the vector fields. The YMF was shown to be renormalizable in Ref.~\cite{tHooft:1972tcz}. The~asymptotic properties of the beta function of Yang--Mills theories, and~particularly of QCD, were studied later, leading to the novel phenomenon of asymptotic freedom~\cite{Politzer:1974fr}. A~comprehensive description of YMF can be found in, e.g.,~Ref.~\cite{tHooft:2005hbu}. 

Fractal is the name given by Benoit Mandelbrot to systems that present a geometry that is very different from the Euclidean one, although it has strong connections with natural phenomena as we observe them. Therefore, fractal geometry has been used to describe many natural shapes that can be observed in everyday life. The~main aspect of a fractal is that it presents a fine structure with an undetermined number of components that are also fractals similar to the original system but at a different scale. This aspect is known as self-similarity. One important characteristic of fractals is the fractal dimension, which~describes how the measurable aspects of the system change with scale. Contrary to the usual quantities in Euclidean geometry, where measurements with higher precision give a better evaluation of the same value for the measured quantity, in~fractals, an increase of the resolution of the measurement will give a different value. This aspect is described by attributing a fractional dimension to the topological dimension of the system. The~classical example is the coastline of an island, which~has different lengths when measured with different precision. 

A direct consequence of the self-similarity and the scaling properties with a fractional dimension is the power-law behavior of distributions observed for fractals. Thus,~probability distributions are described by 

\begin{equation}
 P(x)=A x^{\beta}\,,
\end{equation}
with $A$ being constant, $\beta$ an exponent related to the fractal dimension, and $x$ is any random variable. The impressive ubiquity of fractals in physical systems, but also in mathematical iterative formulas, gives to the fractal theory an importance that can be noticed in many different areas, as~physics, biology, sociology, and engineering. A~nice introductory account of the applications of fractals can be found in Refs.~\cite{Batty,West}, and a comprehensive description of fractal geometry can be found in Ref.~\cite{Mandelbrot}.

The Tsallis statistics is a generalization of the Boltzmann--Gibbs (B-G) statistics, where the entropy is given by 

\begin{equation}
S_q = - k_\B \,\ln_q\,p(x) \,,
\end{equation}
where $p(x)$ is the probability of $x$ to be observed,  $q$ is the entropic index that quantifies how Tsallis entropy departs from the extensive B-G statistics, and $k_\B$ is the Boltzmann constant. In~the above expression, we used the $q$-logarithm function, defined as 
\begin{equation}
\ln_q(p) = \frac{p^{1-q}-1}{1-q}\,. \label{eq:logq}
\end{equation}

A direct consequence of the entropic form defined above is the non-additivity of entropy, since for two independent systems, $A$ and $B$, the entropy of the combined system, $AB$ is
\begin{equation}
S_q(A+B)= S(A) + S(B) + k_\B^{-1}(q-1)S(A)S(B)\,.
\end{equation}

As $q \rightarrow 1$ the B-G statistics is recovered, the entropic form becomes additive and $\ln_q(p) \rightarrow \ln(p)$.

Another consequence is that often the probability distributions obtained in the non-extensive thermodynamics that result from the Tsallis entropic form are described in terms of the $q$-exponential function
\begin{equation}
 e_q(x)=[1+(q-1)x]^\frac{1}{q-1}\,, \label{eq:eq}
\end{equation}
which has found wide applicability in the last few years; see, e.g.,~Refs.~\cite{Tempesta:2011vc,Kalogeropoulos:2014mka}. However, the full understanding of this statistics has not been accomplished yet. A~comprehensive description of Tsallis Statistics and its applications can be found in Refs.~\cite{Tsallis:1987eu,Tsallis:review, Tsallis:book}.

\section{Objectives}
\label{sec:Objectives}

The main objective of this work is to give a short review of the
recent advances in the understanding of the fractal structures present in Yang--Mills fields, in~particular QCD, and~on its main implication, namely the need of Tsallis statistics to describe the thermodynamics aspects of the fields. We will obtain the entropic index of the Tsallis statistic as a function of the \mbox{field fundamental parameters.}

The main features of the fractal structure, which~are summarized by the fractal dimension or fractal spectrum, will be discussed. We will show how the Tsallis statistics entropic index %from the Tsallis 
is related to the fractal structure that is represented as a thermofractal. Then, focusing in QCD and in high energy collisions (HEC), we shall obtain the effective coupling constant in terms of \mbox{the entropic index, $q$,} thereby relating the coupling to the field parameters by an analytical expression. We will show how the power-law behavior observed in momentum distributions measured at high energy collisions is naturally explained by the thermofractal structure. We will mention several experimental observations that can be easily described by the theory cited above. 

The paper is organized as follows: In Section~\ref{sec:Theoretical_Method}, we give a short description of Yang--Mills field theory, emphasizing the renormalization group invariance of the theory and evidencing the fractal structures present in the fields. We also show how the thermodynamic aspects of the Yang--Mills fields reflect the fractal structure in a way similar to the thermofractals, which were introduced to show how Tsallis statistics can emerge in thermodynamic systems, and~study its main thermodynamic quantities. In~Section~\ref{sec:results}, we describe the main experimental findings that give support to the theory presented here, and~we study some applications of the theory to HEP, hadron physics, and astrophysics. Finally, in Section~\ref{sect:Conclusions}, we present a summary of the paper and our conclusions.

\section{Theoretical Method}
\label{sec:Theoretical_Method}
\unskip
\subsection{The Fractal Structures in Yang--Mills Fields} \label{sect:FYMF}

As mentioned above, one of main aspects of the YMF theory, mathematically described by Equation~(\ref{YMLagrangean}), is that it is renormalizable. It~means that vertex functions that are regularized to avoid the ultra-violet (UV) divergence are related to the renormalized vertex functions to which renormalized parameters, $\bar{m}$ and $\bar{g}$, are associated by~\cite{Dyson:1949ha,GellMann:1954fq}
\begin{equation}
\Gamma(p,m,g)=\lambda^{-D} \Gamma(p,\bar{m},\bar{g}) \,,
\end{equation}
where $\Gamma$ is the irreducible truncated vertex function for a particular set of operators, $p$ is the momentum, $m$ the mass, $g$ the coupling constant, $D$ the space-time dimension, and~$\lambda$ is a scale transformation parameter, i.e.,~$p^\mu \to p^{\prime\, \mu} = \lambda p^\mu$.

\begin{figure}[t]%
\centering
 \begin{tabular}{c@{\hspace{3.5em}}c}
 \includegraphics[width=0.37\textwidth]{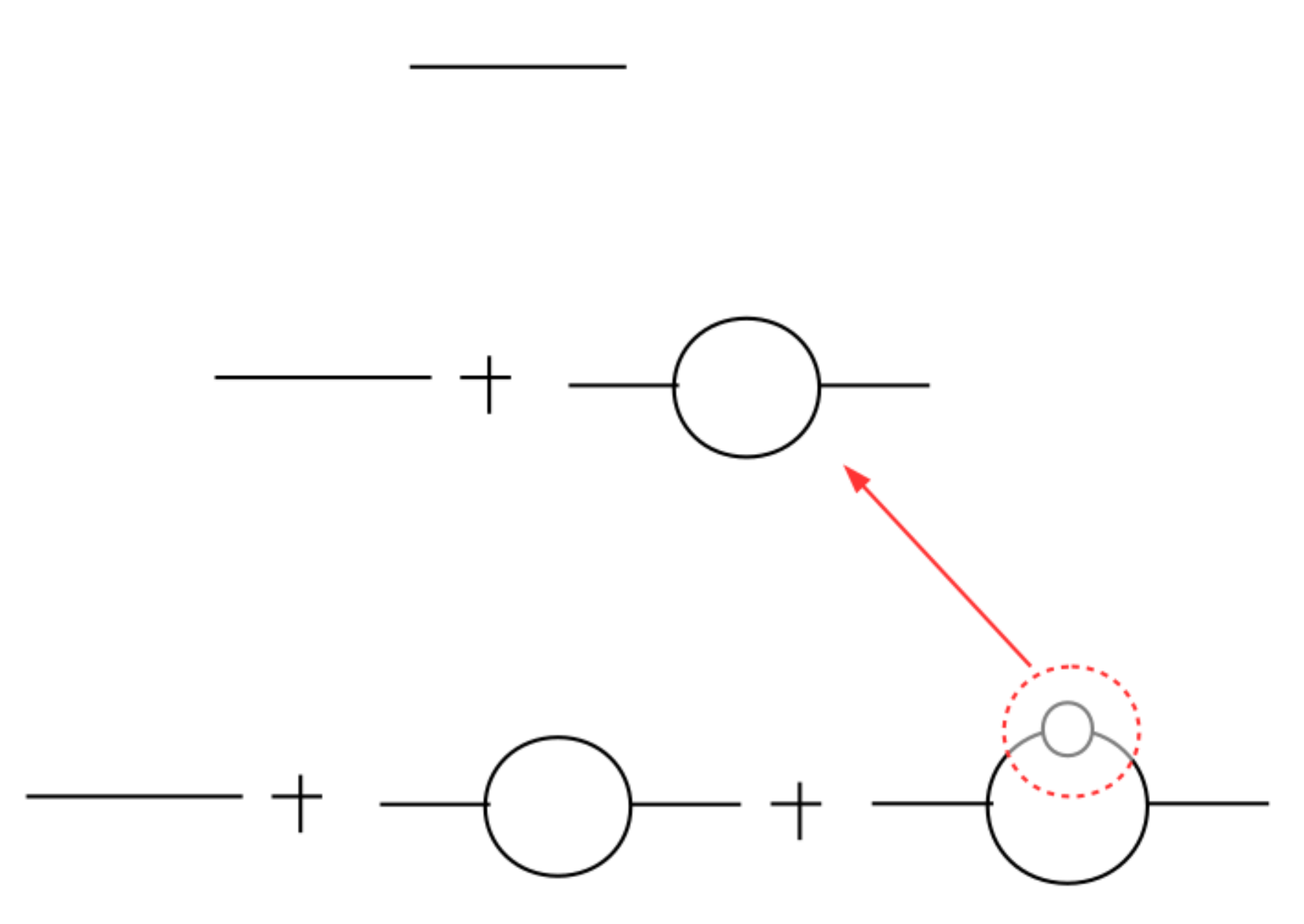} &
\includegraphics[width=0.26\textwidth]{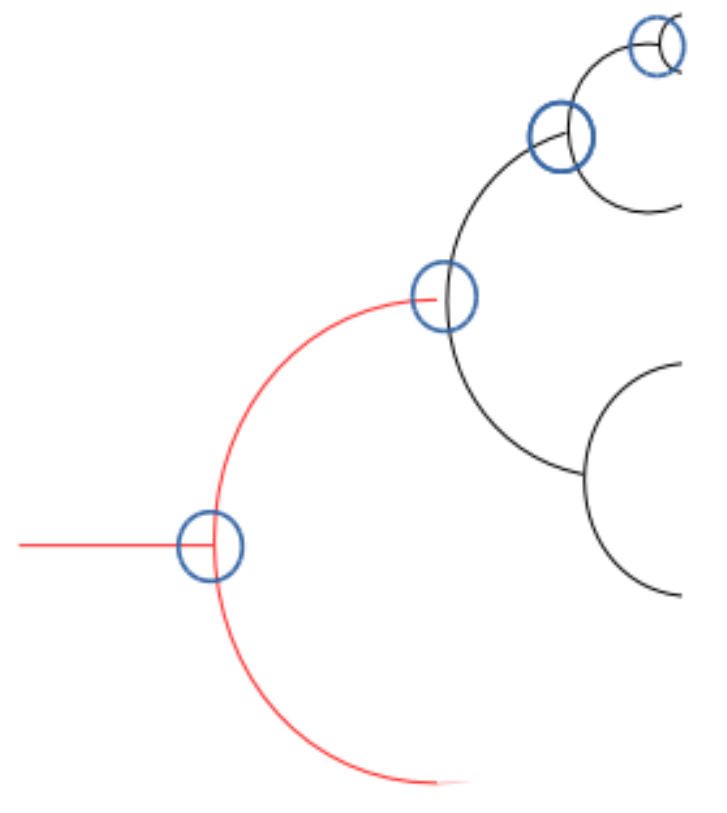} 
\end{tabular}
%\vspace{-4cm}
 \caption{Left panel: Diagrams showing the scaling properties of Yang--Mills fields. It~is shown the loops at different orders in the perturbative expansion. Right panel: Diagrams of multiparticle production in $pp$ collisions. The~initial parton (first black line from the left) may be considered as a constituent of another parton (first red line from the left). 
 The effective vertex couplings, which~are given by the second expression in Equation~(\ref{eq:effective_coupling}) are indicated by circles. The~figure in the left panel is taken from Ref.~\cite{Deppman:2019yno}.
\label{fig:scaling_multiparticle}}
\end{figure}

We denote with a bar the renormalized quantities. This property is described by the renormalization group equation, also known as Callan-Symanzik (C-S) equation, which~is given by~\cite{Callan:1970yg,Symanzik:1970rt}:
\begin{equation}
\left[M\frac{\partial}{\partial M}  + \beta_{\bar g} \frac{\partial}{\partial \bar{g}}  + \bar \gamma \right]\Gamma=0 \,,
\end{equation}
where $M$ is the scale parameter, and~the beta function is defined as $\beta_{\bar g} = M \frac{\partial \bar g}{\partial M}$, and $\bar\gamma$ is the anomalous dimension. As it is shown in the left panel of Figure~\ref{fig:scaling_multiparticle}, renormalization group invariance means that, after proper scaling, the loop in a higher order graph in the perturbative expansion is identical to a loop in lower orders. This is a direct consequence of the C-S equation, and~it is of fundamental importance in what follows. 

The C-S equation is a specific case, for QCD, of a general equation representing renormalization group invariance, and~can be applied to any system where that symmetry is present, thereby all calculations made for the hadronic case apply in a similar form to the Electro-Weak fields. Let us now analyze the possibility of fractal structures to be formed by the Yang--Mills fields.

On the other hand, the scaling symmetry is one of the key ingredients in a fractal system, \mbox{but not the only one}. If we want to show that YMF can form fractal structures, it is important to show that, beside the scaling symmetry it presents an internal structure. The~combination between the scaling property and the internal structure will give rise to the self-similarity, a distinguishing aspect of fractals. We will use the perturbative method and the concept of effective parton, that is, a one-particle irreducible representation of the associated quantum field, to show that the effective parton can be understood as a complex system with internal structure. This internal structure will lead to the formation of fractal structures.

It is clear that the fractal structures may be more easily identified
in the strongly interacting systems governed by QCD, and~in fact,
evidences of the fractal structure can be found in the literature
already a long time ago. In~the 1950’s the existing experimental data
on HEC evidenced the formation of a system in thermodynamic
equilibrium. Indeed, Fermi proposed a thermodynamic model to describe
hadronic collisions at high energies~\cite{Fermi:1950jd}. The~main
issue then was to explain how a short-lived system (half-life of a few
fm/c) could reach equilibrium so fast. This system was called
fireball.

A few years later, Hagedorn~\cite{Hagedorn:1965st} proposed his self-consistent
thermodynamics (SCT) theory that was able to explain several features
of hot hadronic systems. His starting point was a rather weird
definition for fireballs, stating that: 

\vspace{6pt}
{\it ``A fireball is (*) a
  statistical equilibrium (hadronic black-body radiation) of an
  undetermined number of all kinds of fireballs, each o which, in
  turn, is considered to be (goto *)''.}
  \vspace{6pt}

Notice how the definition of fireball given by Hagedorn resembles that of fractals.

Let us introduce in this section the formalism that will allow us to understand the fractal structure of gauge theories. We will consider that at any scale the system can be described as an ideal gas of particles with different masses, i.e masses might change with the scale. The~time evolution of an initial partonic state is given by
\begin{equation}
\ket{\Psi} \equiv \ket{\Psi(t)}=e^{-iHt} \ket{\Psi_0} \,,
\end{equation}
where $H$ is the Hamiltonian of the system. The~state $\ket{\Psi}$ can be written as  $\ket{\Psi}=\sum_{\{n\}} \braket{\Psi_n|\Psi} \ket{\Psi_n}$, where $\ket{\Psi_n}$ is a state with $n$ interactions in the vertex function. Each proper vertex gives rise to a term in the Dyson series, and~at time~$t$ the partonic state is given by
\begin{equation}
 \ket{\Psi} =  \sum_{\{n\}} \frac{(-i)^n}{n!} \int dt_1 \dots dt_n  \, e^{-iH_0 (t_n-t_{n-1})} g  \dots e^{-iH_0 (t_1-t_0)}  \ket{\Psi_0} \,,
\end{equation} 
where $H_0$ is the Hamiltonian with the interaction neglected, $g$ represents the coupling constant, and $t_n>t_{n-1}>\dots>t_1 >t_0$, while $\sum_{\{n\}}$ runs over all possible terms with $n$ interaction vertices. Let us introduce states of well-defined number of effective partons, $\ket{\psi_N}$, so that
\begin{equation}
 \ket{\Psi_n}=\sum_{N} \braket{\psi_N|\Psi_n} \ket{\psi_N}\,. 
\end{equation}

Therefore, $\ket{\psi_N}={\cal S} \ket{\eta_1,m_1,p_1; \dots ; \eta_N,m_N,p_N}$, where ${\cal S}$ is the scattering matrix, while $m_i$ and $p_i$ are the mass and momentum of the $i$ partonic state, and~$\eta_i$ represents all relevant quantum numbers necessary to completely characterize the partonic state. These states can be understood as a quantum gas of particles with different masses. Let us remark at this point that the number of particles in the state $\ket{\Psi_n}$ is not directly related to $n$, since high order contributions to the $N$ particles states can be important. The~rule is $N \le N_{\max}(n) = n(\tilde{N}-1)+1$, where $\tilde{N}$ is the number of particles created or annihilated at each interaction. In~Yang--Mills field theory, $\tilde{N}=2$. We remark that 3- and 4-point contributions are not considered in our approach, since they cannot be renormalized at each order individually, cf. Ref.~\cite{Prochazka:2016ati}.

Let us study the probability to find a state with one parton with mass between $m_0$ and $m_0 + dm_0$, and~momentum between $p_0$ and $p_0 + dp_0$. This is given by
\begin{equation}
\braket{\eta_0,m_0,p_0, \dots|\Psi(t)} = \sum_n  \sum_{N} \braket{\Psi_n|\Psi(t)} \braket{\psi_N|\Psi_n} \braket{\eta_0,m_0,p_0,\dots|\psi_N} \,. \label{eq:probPsi}
\end{equation}

There are three factors in the rhs of this equation. The~first one, $\braket{\Psi_n|\Psi(t)}$, is related to the probability that an effective parton with energy between $E$ and $E + dE$ at time $t=0$ will evolve in such a way that at time $t$ it will generate an arbitrary number of secondary effective partons in a process with $n$ interactions. This factor can be written as $\braket{\Psi_n|\Psi(t)}=G^n P(E) dE$, where $P(E)$ is the probability distribution of the initial particle, and~$G^n$ is the probability that exactly $n$ interactions will occur in the elapsed time. The~second bracket is the probability to get the configuration with $N$ particles after $n$ interactions, i.e.,
\begin{equation}
\braket{\psi_N|\Psi_n}=C_N(n) \stackrel[n \gg 1]{\simeq}{}   \left( \frac{N}{n(\tilde{N}-1)} \right)^4 \,.
\end{equation}

Finally, the last bracket in Equation~(\ref{eq:probPsi}) can be calculated statistically, leading to the following result~\cite{Deppman:2019yno}:
\begin{equation}
\braket{\eta_0,m_0,p_0,\dots|\psi_N} \simeq A(N) P_N\left( \frac{\varepsilon_j}{E} \right)  d^4\left( \frac{p_j}{E} \right) \,,
\end{equation}
with
\begin{equation}
A(N) = \frac{\Gamma(4N)}{8\pi \Gamma(4(N-1))}  \qquad \textrm{and} \qquad P_N\left( x \right) = (1-x)^{4N-5} \,,
\end{equation}
where $\Gamma(x)$ is the Euler Gamma function, $p_j^\mu = (p_j^0, \vec{p}_j)$ is the four-momentum of particle $j$ inside the system of $N$~particles, and~$\varepsilon_j = p_j^0$ is the energy of that particle. Note that we are not assuming a fixed value for the mass $m_j$ of particle $j$, where $m_j^2 = p^\mu p_\mu$, so that $p_j^0$ and $\vec{p}_j$ are variables that may change independently of each other. By combining all these results in Equation~(\ref{eq:probPsi}), one finally obtains 
\begin{equation}
\tilde{P}(\varepsilon_j) d^4p_0 dE \equiv \braket{\eta_0,m_0,
  \dots|\Psi(t)} = \sum_n \sum_{N} G^n
\left(\frac{N}{n(\tilde{N}-1)}\right)^4
\left(1+\frac{\varepsilon_j}{E}\right)^{-(4N-5)} d^4\left(\frac{p
}{E}\right) \left[P(E)\right]^{\nu}dE \,, \label{eq:probPsi_result}
\end{equation}
where $\nu$ is a parameter that, together with the parameter $\tilde{N}$, determine the characteristics of the thermofractal. Its meaning will be elucidated below. We used, in Equation~(\ref{eq:probPsi_result}), that for $N$ sufficiently large and $x \ll 1$, one can approximate $\left( 1 - x  \right)^{(4N-5)}  \simeq \left( 1 + x  \right)^{-(4N-5)}$. Notice that the total energy of the system formed in high energy collisions is of the order of tens of TeV, while the momentum range considered in most experiments is of the order of tens of MeV. Therefore, we can consider $x=(q-1)\varepsilon_j/E \ll 1$, where this identification of $x$ is motivated in Section~\ref{subsec:self_similarity}. 

As we will show below, the distributions $\tilde{P}(\varepsilon)$ and $P(E)$ can be obtained from considerations about self-similarity. Let us mention that relativistic corrections do not play the role of mass variations, as~any relativistic correction should keep the rest mass invariant. Notice that going from one level to the next level of the hierarchy of subsystems does not correspond to a Lorentz transformation, but instead to a scale transformation. Then, the four-momenta~$p^\mu$ do not provide themselves the self-similar properties of the system.

\subsection{Self-Similarity and Fractal Structure}
\label{subsec:self_similarity}

Due to scale invariance, the energy distribution of a parton, as~given by Equation~(\ref{eq:probPsi_result}), \mbox{must depend on the ratio} $\chi = \varepsilon/ E$, and~this ratio must be invariant when we go from one level in the fractal structure to the other. For instance, let us consider that the system with energy $E$, in~which the parton with energy~$\varepsilon_j$ appears as one among $N$ constituents, is itself a parton inside a larger system with energy $\cal{M}$. Then, the scale invariance, when expressed in terms of the variable $\chi$, gives
\begin{eqnarray}
 \frac{\varepsilon_j}{E}=\chi=\frac{E}{{\cal M}}\,.
\end{eqnarray}

Expressing Equation~(\ref{eq:probPsi_result}) in terms of this scale-free variable, we have
\begin{equation}
\tilde{P}(\chi) = \sum_n \sum_{N} G^n
\left(\frac{N}{n(\tilde{N}-1)}\right)^4
\left(1+\chi\right)^{-(4N-5)}  P(\chi) \,. \label{eq:probPsi_result2}
\end{equation}

On the other hand, self-similarity implies that
\begin{equation}
\tilde{P}\left( \chi \right) \propto P\left( \chi \right) \,,
\end{equation}
since the effective partons at any scale have the same probability distribution in terms of the scale-free variable.
This relation imposes a strong condition to the probability distribution in Equation~(\ref{eq:probPsi_result2}), \mbox{since both distributions}, $\tilde{P}(\varepsilon)$ in the left-hand side and $P(E)$ in the right-hand side, must have the same shape. It~is straightforward to conclude that they must follow a power-law function of the form
\begin{equation}
 P(\chi)=(1+\chi)^{\alpha} \,.
\end{equation}

Substituting the ansatz above into Equation~(\ref{eq:probPsi_result2}), it can be shown that~\cite{Deppman:2019yno}
\begin{equation}
P\left( \frac{\varepsilon}{\lambda}\right) = \left[ 1 + (q-1) \frac{\varepsilon}{\lambda} \right]^{-\frac{1}{q-1}} \,, \label{eq:Pe}
\end{equation}
where $q-1 = (1-\nu)/(4N-5)$, while $\nu$ represents the fraction of total number of degrees of freedom (d.o.f.) of the state $\ket{\psi_N}$ that is involved in each interaction, and~$\lambda = (q-1)\Lambda$ is a reduced scale, while $\Lambda$ is the renormalization energy scale of the theory. This probability distribution is a power-law function, and~it will be derived, as well, in Section~\ref{sect:Thermofractals} for thermofractals, cf. Refs.~\cite{Deppman:2016fxs,Deppman:2017fkq,Deppman:2019opw}. 
An interesting interpretation of the entropic index $q$ arises: it is related to the number of internal d.o.f. in the fractal structure, and~the distribution of Equation~(\ref{eq:Pe}) describes how the energy received by the initial parton flows to its internal d.o.f.. In~the context of the theory developed in this section, this probability describes how the energy flows from the initial parton to the partons that appear at higher perturbative orders. Since new orders are associated to new vertices, this suggests that this distribution plays the role of an effective coupling constant in the vertex function, i.e.,
\begin{equation}
 \Gamma=\braket{\Psi_{n+1}| g \, e^{iH_0 t_{n+1}}|\Psi_n} \quad \textrm{with} \quad g=\prod_{i=1}^{\tilde{N}}G\left[1+(q-1)\frac{\varepsilon_i}{\lambda}\right]^{-\frac{1}{q-1}}\,. \label{eq:effective_coupling}
\end{equation}

The situation is schematized in the right panel of Figure~\ref{fig:scaling_multiparticle}, where the vertex functions that are responsible for the scaling properties of the theory in multiparticle production (see Section~\ref{subsec:multiparticle}) are~shown. 

\subsection{Effective Coupling and $\beta$-Function}
\label{sec:beta_function}

The C-S equation together with the renormalized vertex functions were used to derive the beta function of QCD, which~allowed it to show that QCD is asymptotically free~\cite{Politzer:1974fr,Gross:1974cs}. The~result at the one-loop approximation is
\begin{equation}
\beta_{\QCD} = - \frac{\bar{g}^3}{16\pi^2} \left[ \frac{11}{3} c_1 - \frac{4}{3}c_2 \right] \,, \label{eq:betaQCD}
\end{equation}
where $c_1 \delta_{ab} = f_{acd} f_{bcd}$ and $c_2 \delta_{ab} = \Tr\left( T_a T_b\right)$ are directly related to the QCD field parameters, as~described by the Yang--Mills Lagrangian in Equation~(\ref{YMLagrangean}). Quantitatively, the parameters~$c_1$ and $c_2$ are related to the number of colors and flavors by $c_1 = N_c$ and $c_2 = N_f/2$. In~this section, we will study the beta function derived with our ansatz, and~compare with that in QCD. 

\begin{figure}[H]%
\centering
 \begin{tabular}{c@{\hspace{3.5em}}c}
 \includegraphics[width=0.43\textwidth]{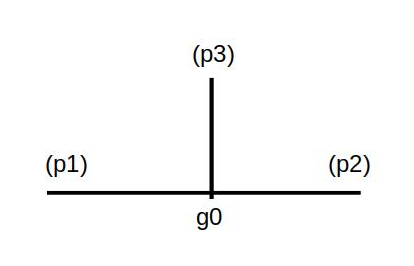} &
\includegraphics[width=0.43\textwidth]{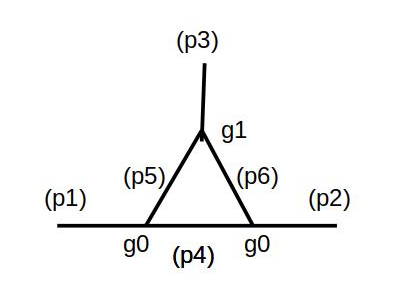} 
\end{tabular}
 \caption{Vertex functions at scale $\lambda_0$ (left) and $\lambda$ (right). Taken from Ref.~\cite{Deppman:2019yno}.}
\label{fig:vertex_functions}
\end{figure}

Let us consider a vertex in two different orders, as~depicted in Figure~\ref{fig:vertex_functions}. The~vertex function at first order, i.e.,~at scale $\lambda_0$, is
\begin{equation}
 \Gamma_o=\braket{\eta_2 p_2 , \eta_3 p_3|g(\lambda_0)e^{iH_0 t}| \eta_1 p_1}\,. \label{eq:initialvertex}
\end{equation}
%%we removed the italic of caption, plz confirm.

The next order in the perturbative approximation is given by the vertex with one additional loop at scale~$\lambda$, which~results in a vertex function
\begin{eqnarray}
\hspace{-1.5cm} \Gamma \hspace{-0.2cm}&=& \hspace{-0.2cm} \bra{\eta_2 p_2 , \eta_3 p_3} g(\lambda_0) e^{iH_0 t_3}\ket{\eta_2 p_6,\eta_3 p_3, \eta_4 p_4} \times  \nonumber \\ 
\hspace{-1.5cm} &&\hspace{-0.2cm}\times  \bra{\eta_2 p_6, \eta_3 p_3, \eta_4 p_4} g(\lambda)e^{iH_0 t_2}  \ket{\eta_1 p_5, \eta_4 p_4} \bra{\eta_1 p_5, \eta_4 p_4}g(\lambda_0)e^{iH_0 t_1} \ket{ \eta_1 p_1}\,. \label{eq:expanded1loopvertexfunction}
\end{eqnarray}

By comparing this expression with $\Gamma=\braket{\eta_2 p_2 , \eta_3 p_3|\bar{g}\, e^{iH_0 t}| \eta_1 p_1}$, one can identify the effective coupling $\bar{g}$ as
\begin{equation}
 \bar{g}=g(\lambda_0) e^{iH_0 t_3}\ket{\eta_2 p_6,\eta_3 p_3, \eta_4 p_4} \Gamma_M \bra{\eta_1 p_5, \eta_4 p_4}g(\lambda_0)\,, \label{eq:1loopcoupling}
\end{equation}
where
\begin{equation}
 \Gamma_M=\braket{\eta_2 p_6, \eta_3 p_3, \eta_4 p_4|g(\lambda)e^{iH_0 t_2} | \eta_1 p_5, \eta_4 p_4}\,. \label{eq:scaledvertex}
\end{equation} 

The indexes $1,2,\cdots,6$ refer to each of the particles created or absorbed during the process as described in the Feynman diagram in Figure~\ref{fig:vertex_functions}. The~symbol $\eta_i$ represents the set of quantum numbers necessary to fully characterize the particles, as~spin, color, and flavor. The~scaling properties of Yang--Mills fields allow us to relate $\Gamma_M$ to $\Gamma_o$ by an appropriate scale,~$\lambda$. From dimensional analysis, the scaling behavior turns out to be $\Gamma_M(\lambda) = (\lambda/\lambda_0)^4$. Using these considerations and from the C-S equation, it results that 

\begin{equation}
\beta_{\bar{g}}\frac{\partial \Gamma}{\partial g}=-(d+\bar\gamma_5+\bar\gamma_6)\Gamma\,,
\end{equation}
where $d=4$ and $\bar\gamma_{5,6}$ are the anomalous dimensions. In~order to compare with the QCD results, we will study the behavior of $g(\lambda)$ at $\lambda = \lambda_o/\mu$, where $\mu$ is a scaling factor. From Equation~(\ref{eq:effective_coupling}), one has

\begin{equation}
 g(\mu)=\prod_{i=5}^{6}G\left[1+(q-1)\frac{\varepsilon_i \mu}{\lambda_o}\right]^{-\frac{1}{q-1}}\,, \label{eq:QCDrunningcoupling}
\end{equation}
and, substituting it into Equations~(\ref{eq:1loopcoupling}) and (\ref{eq:scaledvertex}), one can calculate the beta function in the one loop approximation. By considering the asymptotic limit $(q-1)\mu \gg \lambda_0 / \varepsilon_i$,
one obtains
\begin{equation}
\beta_{\bar{g}}=-\frac{1}{16\pi^2} \frac{1}{q-1} \bar{g}^{\tilde{N}+1}\,, \label{eq:betag}
\end{equation}
with $\tilde{N} =2$. Finally, by comparing with the QCD result, Equation~(\ref{eq:betaQCD}), one can make the identification
\begin{equation}
\frac{1}{q-1} = \frac{11}{3}c_1 - \frac{4}{3}c_2 = 7 \,,
\end{equation}
where, in the last equality, we used $N_c = N_f/2 = 3$. This result
leads to $q = 1.14$, leading to an excellent agreement with the
current experimental data analyses, as~it will be discussed in
Section~\ref{sec:results}. Notice that this analysis also predicts that
there is some dependence of $q$ on the number of flavors. It~would be
interesting that this prediction being confronted with experimental
$q$-fits studies.

We notice that the scaling properties used here are necessary conditions to allow the renormalization of the quantum field theory after regularization and~are present {\it independently} of the renormalization scale used; hence, our results are valid at any value for the scale chosen. \mbox{In~our approach}, it becomes evident that the number of flavors gives the scaling dimension of the theory, and~it is an essential part of the renormalization procedure, no matter what is the scale used to fix the theory after regularization.

\begin{figure}[H]%
\centering
 \begin{tabular}{c@{\hspace{3.5em}}c}
 \includegraphics[width=0.4\textwidth]{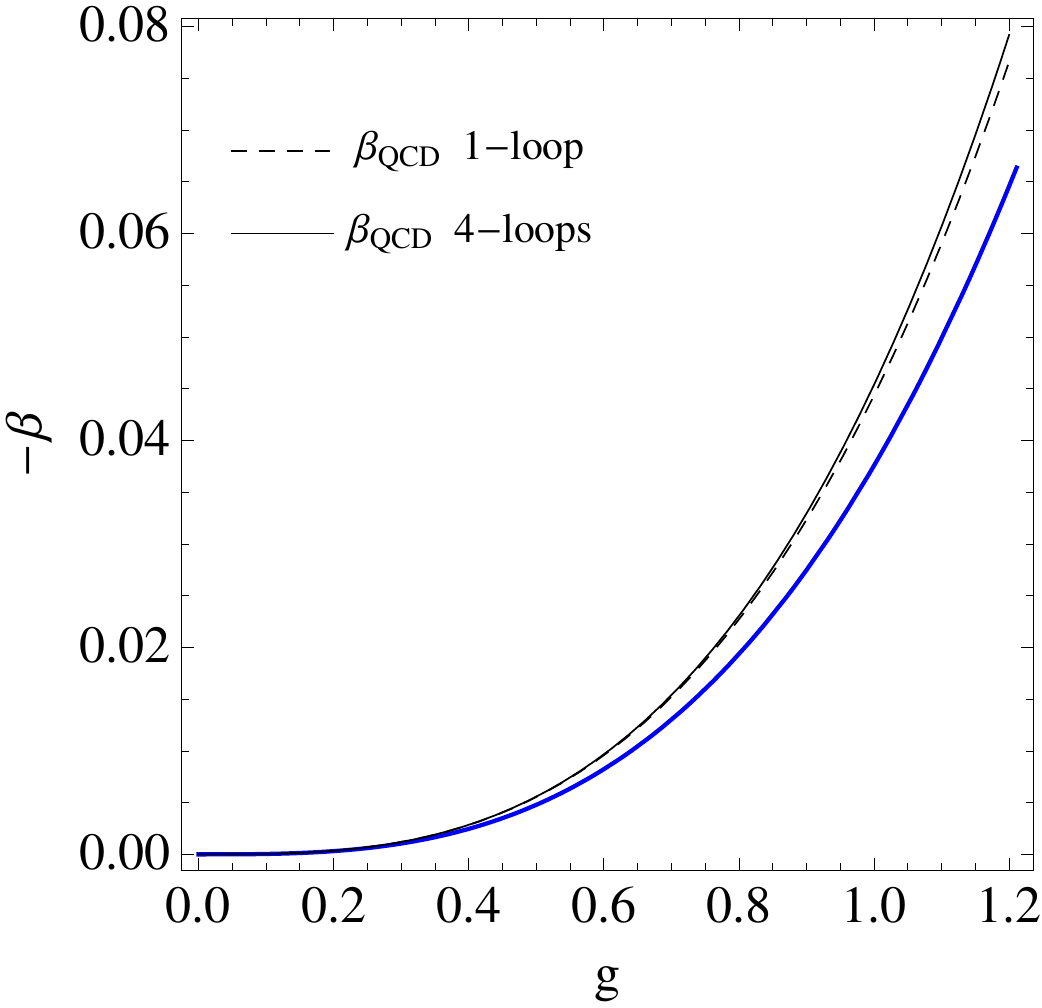} &
\includegraphics[width=0.4\textwidth]{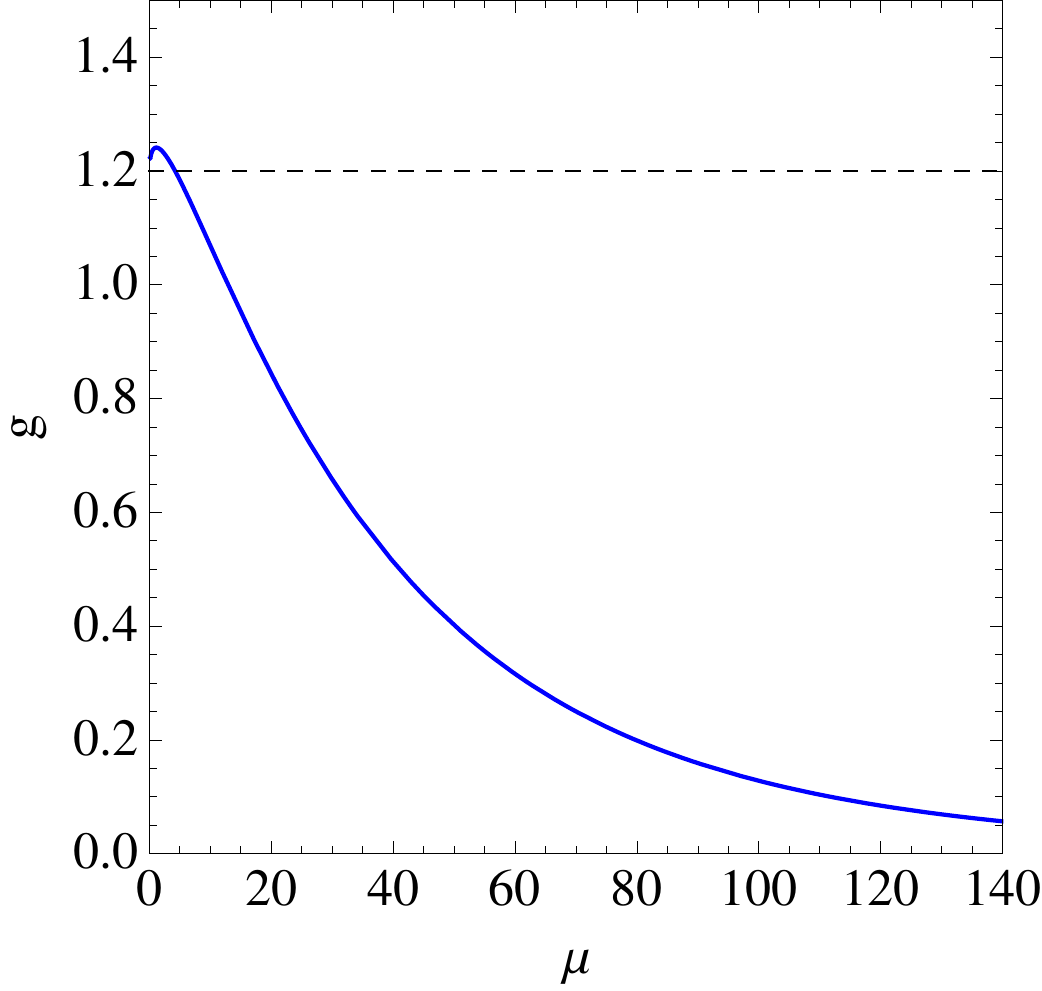}
\end{tabular}
 \caption{Left panel: Behavior of beta function against effective coupling, as~given by Equation~(\ref{eq:betag}). We display also the result from Quantum
Chromodynamics (QCD) in one-loop~\cite{Politzer:1974fr} and four-loop~\cite{Czakon:2004bu} approximation. Right panel: Effective coupling as a function of the scale~$\mu$, as~calculated by Equation~(\ref{eq:QCDrunningcoupling}). Taken from Ref.~\cite{Deppman:2019yno}.}
\label{fig:betag}
\end{figure}

In Figure~\ref{fig:betag}, we display the plots showing the behaviors of the beta function $\beta_{\bar{g}}$ as a function of $g$, and~of the coupling $g$ as a function of~$\mu$. The~results obtained here give a strong basis for the interpretation of previous experimental and phenomenological studies on QCD in terms of non-extensive statistics and thermofractals. 
%%we removed the italic of caption, plz confirm.

Finally, one can see from the result of Equation~(\ref{eq:betag}) that values of $q > 1 \, (< 1)$ lead to negative(positive) values of the beta function, i.e.,~$\beta_{\bar{g}} < 0 \, (>0)$. While we will mainly focus on the case $q > 1$ in this manuscript, values $q < 1$ are also allowed by the present description, and~this corresponds to a YMF with a positive beta function as it follows from Equation~(\ref{eq:betag}); this is the case of quantum electrodynamics (QED). This illustrates the fact that the present formalism might have an impact not only in QCD physics but also in research fields like atomic and condensed matter physics.

\subsection{Thermofractals}
 \label{sect:Thermofractals}

So far, we have shown that, in~principle, fractal structures are possible to be formed in YMF. \mbox{What we have to }consider now is the evidence that such structures are really present in the physical systems, and if they play a relevant role in the evolution of those systems in time and in the interactions with other fields. It~is easy to observe that evidence of such fractal structure was noticed \mbox{long time ago.}

The Hagedorn's SCT gave predictions for some quantities that could be easily accessed experimentally. In~particular, it predicted the transverse momentum distribution of the particles produced in the decay of the fireball, given by

\begin{equation}
\frac{d {\mathcal N}}{dp_\perp} \bigg|_{y=0} = g V \frac{p_\perp m_\perp}{(2\pi)^2} \exp\left( \frac{m_\perp}{T} \right) \,, \label{eq:dNdp_BG}
\end{equation}
where $g$ is a constant, $V$ is the volume of the system, $m_\perp = \sqrt{p_\perp^2 + m^2}$, and $y$ is the rapidity. Another prediction was the hadron mass spectrum, given by
\begin{equation}
\rho(m) = \gamma m^{-5/2} \exp\left( \beta_0 m \right) \,, \label{eq:rho_m}
\end{equation}
where $\beta_0= 1/T_0$, while $T_0$ and $\gamma$ are the parameters.

One interesting aspect of the theory was the prediction of a limiting
temperature for the fireball, known today as Hagedorn temperature,
$T_{\rm H}$, which~is numerically equal to the parameter $T_0$, \mbox{i.e.,~$T_{\rm H} =
T_0$.} The comparison between theory and experiment brought a sudden
recognition of the importance of Hagedorn’s theory, and~its impact in
HEP is notorious. Frautschi proposed a similar theory-based
exclusively in hadron structure, stating that ``hadrons are made of
hadrons'', again using self-reference~\cite{Frautschi:1971ij}. With
this so-called bootstrap model, he obtained the same hadron mass
spectrum formula shown in Equation~(\ref{eq:rho_m}). 

The success of SCT prompted the development of ideal gas models for
hadronic systems, \mbox{the hadron resonance} gas (HRG)
models~\cite{Hagedorn:1984hz,Yukalov:1997jk,Cleymans:1999st,Agasian:2001bj,Tawfik:2004sw,Megias:2009mp,Huovinen:2009yb,Borsanyi:2010cj,Bazavov:2011nk,Megias:2012kb,NoronhaHostler:2012ug},
that were able to explain many features of high energy collisions and
hadron physics. Before that, an important
paper~\cite{Dashen:1969ep} had explained, through Dyson--Schwinger
expansion, how a strongly interacting system could behave under some
conditions as an ideal gas of resonant particles. %MDPI: please try to
%avoid names, simply write "an important paper [37]". You do NOT use to
%MANY other refs. the names, so it look strange.

Another important consequence of SCT came with the advent of the quark structure of hadrons, that was used in Ref.~\cite{Cabibbo:1975ig}  %MDPI: again please remove the names as soon as thpose are not the namels of formula, approach etc. (like e.g. the Dyson-Schwinger expansion just a couple of lines above), so it reads: (lines 6-7): "that was used in Ref. [38], to"
, to
propose that Hagedorn temperature was not a limiting temperature but a
transition temperature between the confined quark and the deconfined
quark regimes of hadronic matter. The~last phase is known as
QGP and is one of the most interesting issues in
nowadays nuclear physics. Despite its initial success and its
important consequences, with the results coming from accelerators able
to deliver particles at higher energies than those from existing at
that time, it was found that SCT was not able to explain correctly the
outcome of the new HEP data. Indeed, Hagedorn himself proposed, in
substitution to this thermodynamics theory, an empirical model based
on QCD~\cite{Hagedorn:1983wk}. \mbox{Since his theory gave so much} correct
information about hadronic systems, the question imposes itself is:
what went wrong with Hagedorn’s theory?

A possible answer to this question is based on the special fractal
structure found for YMF, and~hence to QCD, which~implies in the non-extensive
thermodynamics that hadronic systems, and~supposedly hadron, may present~\cite{Deppman:2016fxs}. Such a fractal
structure would result in fractional dimension of the phase-space of
the compound particles of the system. Preliminary analysis of such
idea lead to a fractal dimension that is compatible with the results
from intermittency analyses.

The self-similarity is an evident feature of fireballs and hadrons
according to their definitions by Hagedorn and Frautschi,
respectively, as~described above. However, many other evidences point
to the self-similar structure of hadrons: fractal dimension has been
identified through the analysis of intermittence in experimental
distributions obtained in HEP experiments~\cite{Bialas:1985jb,Bialas:1988wc,Hwa:1989vn,Hwa:1991rd,Hegyi:1992dj,Hegyi:1993ia,Dremin:1993ee,Sarkisyan, Kittel,DeWolf:1995nyp}; a parton distribution function (PDF) that describes the proton structure based on fractal properties was shown to fit experimental
data rather well~\cite{Lastovicka:2002hw}; direct evidences of self-similarity in experimental data has been observed~\cite{Wilk:2009nn}, but the so-called z-scaling~\cite{Tokarev:2020wyr,Tokarev:2020ujg} might be related to fractal structures, as well. 

The emergence of the non-extensive behavior has been attributed to
different causes. These include long-range interactions, correlations and memory
effects~\cite{Borland:1998}, temperature fluctuations, and~finite size of the
system, among others. We will show in this section that a natural derivation of non-extensive statistics in terms of thermofractals~\cite{Deppman:2019yno, Deppman:2016fxs} is possible. These are systems in thermodynamical equilibrium presenting the following properties:

\begin{enumerate}[leftmargin=*,labelsep=4.9mm]
\item Total energy is given by~$U = F + E$, where $F$ is the kinetic energy, and~$E$ is the internal energy of $N$ constituent subsystems, so that $E = \sum_{i=1}^N \varepsilon_i$. 
\item The constituent particles are thermofractals. This means that the energy distribution $P_{\TF}(E)$ is self-similar or self-affine, i.e.,~at level $n$ of the hierarchy of subsystems, $P_{\TF (n)}(\chi)$ is equal to the distribution in any other level, with $\chi$ being a scale-free variable that can be given by $\chi=E/F$ in the case of thermofractals of the type-I, or $\chi=E/U$ for thermofractals of the type-II. 
\item At some level of the fractal structure, the internal energy fluctuation is small enough to be disregarded. In~this case, the internal energy is considered to be equal to the component mass $m$.
\end{enumerate}

We denote by $P_{\rm{B-G}}$ and $P_\TF$ the B-G and thermofractal distributions, respectively. It is possible to show that thermofractals results in energy distributions of the kind
\begin{eqnarray}
P(\varepsilon)=\left[1 \pm (q-1) \frac{\varepsilon}{\lambda} \right]^{\mp1/(q-1)}\,,
\end{eqnarray}
with the negative sign in the exponent corresponding to the type-II thermofractal, while the positive sign corresponds to the type-I. The~main difference between thermofractals of type-I and of type-II is the character of the distribution: type-I presents a distribution without cut-off, while type-II requires a cut-off because of the negative sing in the argument of the $q$-exponential function. %Each type of thermofractal leads to a distribution without cut-off (type-I) or with cut-off (type-II). 
As we will see below, $q$ is the entropic index of the Tsallis statistics, while $\lambda$ is a scale variable. In~what follows, we will consider only the thermofractals of type-I, but the type-II can be derived in the same way~\cite{Deppman:2019yno}.
The energy distribution according to B-G statistics is given by 

\begin{equation}
 P_{\rm{B-G}}(U) dU=A \exp(-U/(k_\B T)) \, dU \,, \label{eq:P_BG}
\end{equation}
where $A$ is a normalization constant. In~the case of thermofractals, the phase space must include the momentum d.o.f. 
$(\propto f(F))$ of free particles, as well as internal d.o.f. $(\propto f(\varepsilon))$. Then, the internal energy is $dE \propto [P_{\TF}(\varepsilon)]^\kappa d\varepsilon$ where $\kappa$ is an exponent to be determined, and~one has~\cite{Deppman:2016fxs}
\begin{equation}
  P_{\TF (0)}(U) dU = A^\prime  F^{\frac{3N}{2}-1} \exp\left( -\frac{\alpha F}{k_\B T} \right) dF \left[ P_{\TF (1)}(\varepsilon) \right]^\kappa d\varepsilon \,,  \label{eq:PTF01_1}
\end{equation} 
with $\alpha = 1 + \varepsilon/(k_\B T)$ and $\varepsilon/(k_\B T) = E/F$. 
This expression relates the distributions at level $0$ and $1$ of the subsystem hierarchy. After integration, and~by imposing self-similarity, i.e.,
\begin{equation}
P_{\TF (0)}(U) \propto P_{\TF (1)}(\varepsilon) \,, \label{eq:PTF01_2}
\end{equation}
 the simultaneous solution of Equations~(\ref{eq:PTF01_1}) and (\ref{eq:PTF01_2}) is obtained with 
  ~\cite{Deppman:2017fkq}
\begin{equation}
 P_{\TF (n)}(\varepsilon)= A_{(n)} \cdot \left[1 + (q-1) \frac{\varepsilon}{k_\B\tau} \right]^{-\frac{1}{q-1}}  = A_{(n)} \cdot e_q\left(- \varepsilon/(k_\B\tau) \right) \,. \label{eq:selfsimilar}
\end{equation}

We find that the distribution of thermofractals obeys Tsallis statistics with $q-1 = 2(1-\kappa)/(3N)$ and $\tau = N (q-1) T$. %This result shows that there is a clear connection between hadron structure, which~is described by Tsallis statistics, and~thermofractals. 
Observe that $q$ is related to the number of d.o.f. that are relevant to the description of the system, and~the fact that $q$ is different from unit means that this number is finite. In~this aspect, thermofractals can be considered as examples of small systems with finite d.o.f.. In~the rest of this manuscript, we will study the connection between thermofractals and quantum field theory, and~in particular to QCD.

\subsection{Non-Extensive Self-Consistent Thermodynamics}

The non-extensive self-consistent thermodynamics (NESCT) is a generalization of the SCT theory by imposing the
self-consistency principle from Hagedorn in the non-extensive
statistics from Tsallis~\cite{Deppman:2012us}. The~basic ingredients
are the two forms of partition function for fireballs proposed by
Hagedorn, only this time using the Tsallis $q$-exponential factor in Equation~(\ref{eq:selfsimilar}), i.e.,
\begin{equation}
P(\varepsilon) = A \cdot e_q\left( - \varepsilon/(k_\B \tau) \right)  \,,\label{eq:P_epsilon}
\end{equation}
which leads to
\begin{equation}
Z_q (V_0,T) = \int_0^\infty \sigma(E) \left[ 1 + (q - 1)\beta E \right]^{-\frac{q}{(q-1)}} dE \,,
\end{equation}
where $\beta = 1/(k_\B \tau)$, and~$\sigma(E)$ is the density of states of the fireballs as a function of its energy, and
\begin{equation}
\log\left[ 1 + Z_q(V_0,T) \right] = \frac{V_0}{2\pi^2} \sum_{n=1}^\infty \frac{1}{n} \int_0^\infty dm \int_0^\infty dp \, p^2 \rho(n;m)  \left[ 1 + (q-1)\beta \sqrt{p^2 + m^2} \right]^{-\frac{nq}{(q-1)}} \,,
\end{equation}
which is the partition function for an ideal gas of fermions and
bosons with mass~$m$. The~sum in $n$ results from the expansion of the
logarithm function, and~$\rho(n; m) = \rho_f(m) - (-1)^n \rho_b(m)$ is
related to the fermionic and bosonic mass spectra. In~the case $n = 1$,
which will be used here, $\rho(1; m) = \rho_f(m) + \rho_b(m) \equiv \rho(m)$ is the hadron mass spectrum. 

Since fireballs are thermofractals and present self-similarity, the
following weak-constraint \mbox{must hold}:
\begin{equation}
\log[\rho(m)] = \log[\sigma(E)] \,. 
\end{equation}

The main task now is to find functions $\rho(m)$ and $\sigma(E)$ that
satisfy the equality above and simultaneously let the two forms of
partition function identical to each other. This can be
achieved~\cite{Deppman:2012us,Megias:2015fra} by choosing 
\begin{equation}
\rho(m) = \frac{\gamma}{m^{5/2}} \left[ 1 + (q - 1)\beta_0 m \right]^{\frac{1}{q - 1}} \,,  \label{eq:rho_m_2}
\end{equation}
and 
\begin{equation}
\sigma(E) = b E^c \left[ 1 + (q - 1) \beta_0 m   \right]^{\frac{1}{q - 1}} \,, \label{eq:sigma_E}
\end{equation}
where $m$ is the mass of the components of the thermofractal and $E$ is the total energy of the system. With those functions both forms of partition function result in
\begin{equation}
Z_q(V_0,T) \rightarrow b \Gamma(c + 1) \left( \frac{1}{\beta - \beta_0} \right)^{a+1}  \,, \label{eq:Zq_sing}
\end{equation}
with 
\begin{equation}
c + 1 = \alpha = \frac{\gamma V_0}{ 2\pi^2 \beta^{3/2}} \,. 
\end{equation}

From here, one can see that there is a singularity at $\beta = \beta_0$, and,~therefore, the limiting temperature is found. The~singularity at $\beta_0$ has the same nature as the limiting temperature in Hagedorn's self-consistent thermodynamics~\cite{Hagedorn:1965st}, and~indicates a limiting temperature for hadronic systems, also known as Hagedorn's temperature. This limiting temperature was interpreted by Cabbibo and Parisi~\cite{Cabibbo:1975ig} as the critical temperature for the phase-transition between confined and deconfined regimes of quark matter.

 Therefore, it is possible to generalize Hagedorn’s self-consistency
 principle within Tsallis statistics and as a result one gets a new
 hadron mass spectrum formula, given by Equation~(\ref{eq:rho_m_2}),
 the limiting temperature resulting from the singularity in
 Equation~(\ref{eq:Zq_sing}), now expressed in terms of the Tsallis
 temperature, but also a universal entropic index, $q$, which~is
 characteristics of any hadronic system. The
   universality of hadroproduction processes in HEP implies that the
   value of $q$ is the same disregard the colliding objects
   (leptons to hadrons to nuclei), a feature already stressed,
   cf. e.g. Refs.~\cite{Sarkisyan:2004vq,Sarkisyan:2005rt}, and widely discussed~\cite{Zyla:2020zbs}. It~is
 worthy of mention that both $T$ and $q$ are parameters in the hadron
 mass spectrum; therefore, they are related to the hadronic structure.

As in the case of the Hagedorn’s theory, cf. Equation~(\ref{eq:dNdp_BG}), here also, the way to test
experimentally the predictions of the theory is through the transversal
momentum distribution, given by 

\begin{equation}
\frac{d{\mathcal N}}{dp_\perp} \bigg|_{y=0} = g V \frac{p_\perp m_\perp}{(2\pi)^2} e_q\left( \frac{m_\perp}{T} \right) \,. \label{eq:dNdpT}
\end{equation}

\begin{figure}[H]%
\centering
 \begin{tabular}{c@{\hspace{3.5em}}c}
 \includegraphics[width=0.46\textwidth]{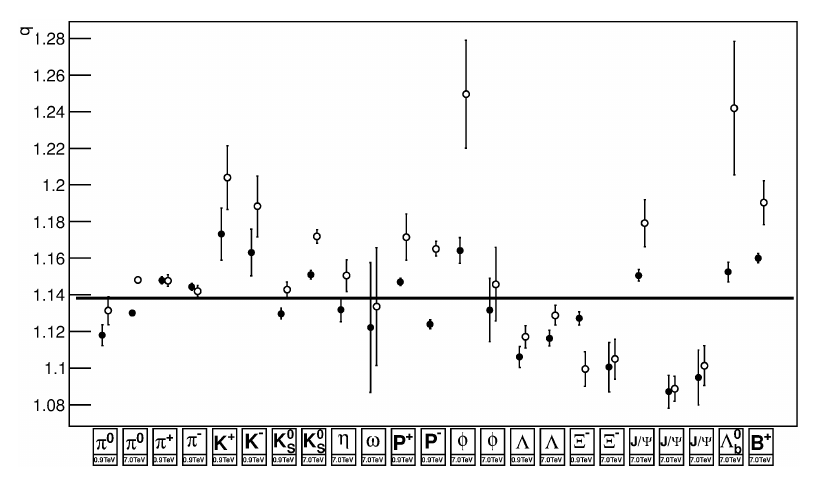} &
\includegraphics[width=0.42\textwidth]{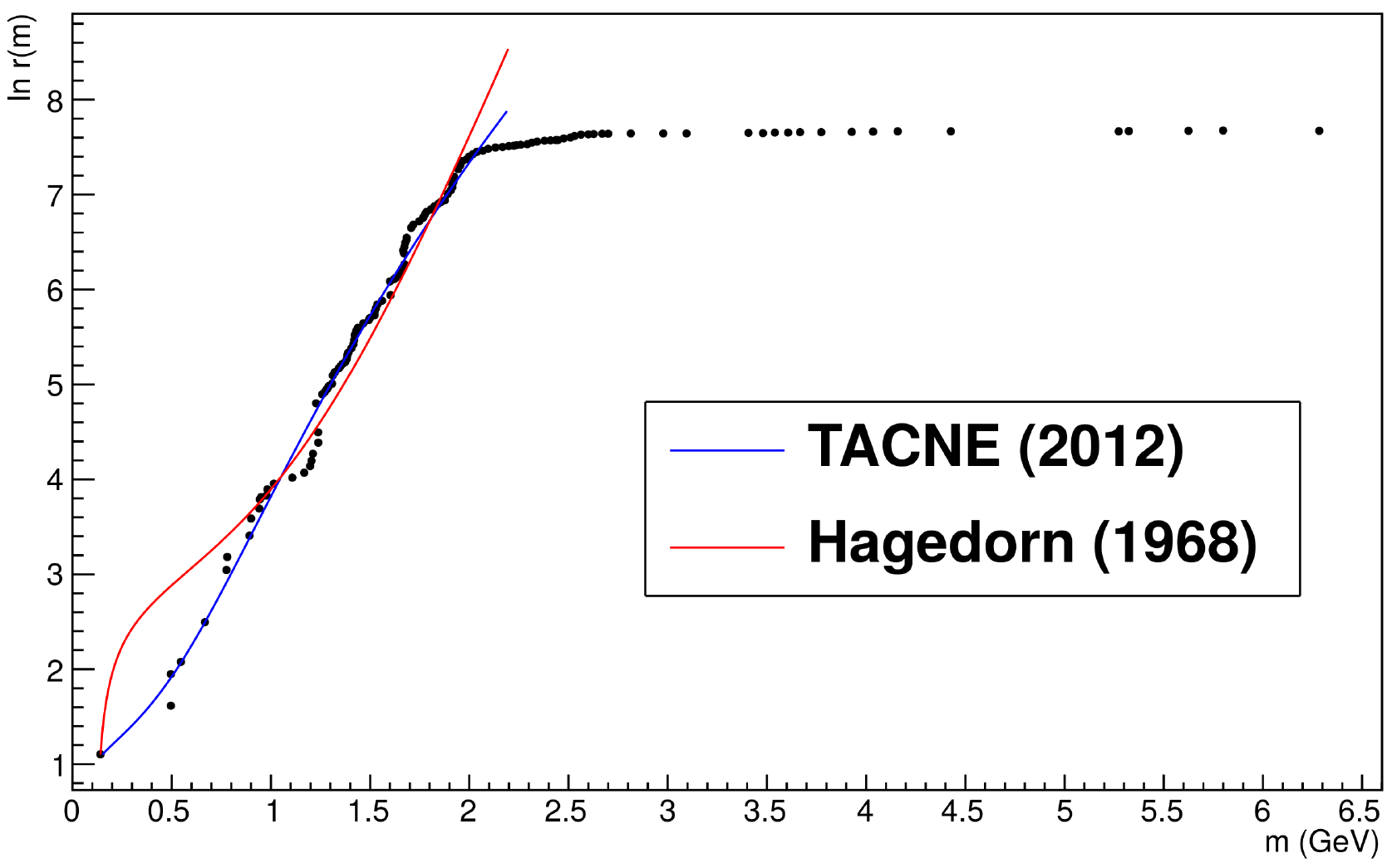}
\end{tabular}
 \caption{Left panel: Entropic index, $q$, obtained from a fit of the abundance of different hadron species in $pp$ collisions using Tsallis statistics, cf. Refs.~\cite{Marques:2012px,Marques:2015mwa}. Right panel: Cumulative hadron spectrum, \mbox{as~a function of} the hadron mass. The~dots stand for the PDG result~\cite{Zyla:2020zbs}. We display also he result by using the non-extensive self-consistent thermodynamics (blue) and the result predicted by Hagedorn (red), \mbox{cf. Ref.~\cite{Marques:2012px}}.} 
\label{fig:experiment}
\end{figure}

Once the parameters $T$ and $q$ are determined by fitting of
Equation~(\ref{eq:dNdpT}) to data on $p_\perp$ distribution, the entire
thermodynamics of hot hadronic systems is determined. The~extension of
the Hagedorn theory to non-extensive statistics, which~is
characterized by the probability distribution
Equation~(\ref{eq:P_epsilon}), allowed to reproduce the distribution of all
the species produced in $pp$ collisions with a high accuracy, leading
to the result~\cite{Marques:2012px,Marques:2015mwa}
\begin{equation}
q = 1.14 \pm 0.01 \,,
\end{equation}
(see left panel of Figure~\ref{fig:experiment}).Other reactions 
like $AA$, $dA$, or $pA$ collisions show the same behavior~\cite{Zyla:2020zbs} (see, e.g.,~Ref.~\cite{Biro:2020kve}, for a recent review). Moreover, the power-law behavior for the hadron spectrum predicted by this extension, cf. Equation~(\ref{eq:rho_m_2}), can be compared with the hadron spectrum from the Particle Data Group (PDG)~\cite{Zyla:2020zbs}. This comparison, as~displayed in the right panel of Figure~\ref{fig:experiment}, leads to an important improvement with respect to the exponential distribution $\rho(m) = \rho_o \, e^{m/T_H}$ proposed by Hagedorn, specially at the lowest masses, cf. Ref.~\cite{Marques:2012px}.~Observe that the hadron spectrum given by Equation~(\ref{eq:rho_m_2}) is valid up to $m=2.5 \, \GeV$, and~we consider $\rho(m)=0$ above this limit. Similarly the Regge trajectories may relax from the straight lines (see Ref.~\cite{Fazio:2011ex} and references therein)). Notice that it would be possible to consider an approximation in the extension of the Hagedorn theory to non-extensive statistics, by using that the product of the functions in Equations~(\ref{eq:P_epsilon}) and (\ref{eq:rho_m_2}) in the partition function can be expressed as

\begin{equation}
\rho(m) P(\varepsilon)  \approx \rho_0 A \left[1+(q-1) \frac{\varepsilon - m}{M} \right]^{-\frac{1}{q-1}} \,.
\end{equation}

This leads to a good fit of observables as long as $m \ll M/(q-1)$.

Finally, let us mention that it is possible to obtain any
thermodynamics quantity and compare it to Lattice QCD (LQCD)
results. A~comparison between the results from NESCT and LQCD data has
shown also a good agreement between both
calculations~\cite{Deppman:2012qt}, in~this case without any
adjustable parameter. Apart from these considerations, there are many
other applications in which Tsallis statistics play an important
role. This includes high energy
collisions~\cite{Cleymans:2011in,Marques:2015mwa,Wong:2015mba,Rybczynski:2020}, hadron
models~\cite{Cardoso:2017pbu}, hadron mass
spectrum~\cite{Marques:2012px}, neutron stars~\cite{Menezes:2014wqa},
LQCD~\cite{Deppman:2012qt}, non-extensive
statistics~\cite{Deppman:2012us,Megias:2015fra,Deppman:2017fkq}, and
many others. In~the rest of the manuscript, we will see that the
power-law behavior of Tsallis statistics leads, in~fact, to important
phenomenological consequences in these fields.

The theory presented here can be extended to $AA$ collisions, as~long as collective motion of the QGP fluid is taken into account~\cite{Grigoryan:2017gcg}.
The analyses of experimental data from HEP give room to some small variation of the entropic index, $q$, and~for the temperature, $\tau$, with collision energy and/or particle species~\cite{Bhattacharyya:2017hdc}, and~the variation is larger for small multiplicity events. The~variation can, in~principle, be due to non-equilibrium processes, where perturbative QCD should be applied. In~fact, as~the multiplicity increases, the results seem to be independent from collision energy, and~remain valid for $pp$, $pA$, and $AA$ collision~\cite{Sharma:2018jqf}. The~use of non-extensive thermodynamics is important for the precise determination of the character of the phase transition between confined and deconfined states of the QCD matter~\cite{Azmi:2019irb,Biro:2017arf}.

We mentioned above that $q$ and $\nu$ are related to the number of d.o.f. in the system. In~the context of the fractal structure exhibited by the YMF, $q$ is related to the number of relevant d.o.f. involved in the transfer of energy and momentum between interacting systems. Thus,~regarding the general applicability of the theory, we observe that the most complex case was considered here, i.e.,~the case where the full phase-space is occupied and all possible field configurations, for both fermions and bosons, are allowed. These conditions seem to be valid for QGP, but for simpler systems, where the total quantum numbers are restricted enough or when topology does not allow the complete phase space to be occupied, as~it may happen in, \mbox{e.g.,~deep inelastic} scattering or in high energy particles scattering, the $q$ value may change accordingly. Recent experimental data show, indeed, that for high multiplicity events the $q$ value tends to a universal value in good agreement with the one found here~\cite{Acharya:2019mzb}.  This universality is also manifest when considering other effects like e.g. eccentricity and flows~\cite{Castorina:2019vex}.

\subsection{Multiparticle Production}
\label{subsec:multiparticle}

An interesting physical situation in which the scaling properties emerge is in multiparticle production in $pp$ collisions~\cite{Konishi:1978ks}. We show in the right panel of Figure~\ref{fig:scaling_multiparticle} a typical diagram in which two partons are created in each vertex. When effective masses and charges are used, the line of the respective field in Feynman diagrams represents an effective particle. An irreducible graph represents an effective parton, and~the vertices are related to the creation of an effective parton. Due to the complexity of the system, it is desirable a statistical description in which the summation of all the diagrams becomes equivalent to an ideal gas of particles with different masses~\cite{Dashen:1969ep,Venugopalan:1992hy}. \mbox{Another example} is given by the hadron structure: As in the case of multiparticle production, too many complex graphs should be considered when studying the hadron at a high resolution. This is why calculations are typically limited either to the first leading orders, or to LQCD methods.

\subsection{Non-Extensive Thermodynamics of Thermofractals}
\label{sec:Tsallis_Statistics}

As we have seen above, the properties of the YMF can lead to the
formation of fractal structures, and~the thermofractal systems show us
that the thermodynamics formalism that describes YMF fractal structure
is the Tsallis statistics. In~QCD, such fractal aspects appear in the
generalized version of Hagedorn's thermodynamics, which~was described
by the NESCT approach. In this section, we will describe some of the
main aspects of the thermofractal system, since most of its
characteristics are important for applications in hadron models and on
the description of massive objects, as~neutron stars, through the
Tsallis statistics. We will present here the thermodynamics of a free
quantum gas of bosons and fermions at finite temperature and chemical
potential within Tsallis statistics.

We will define two versions of the $q$-exponential and $q$-logarithm functions as follows:
\begin{equation}
e_q^{(\pm)}(x)=[1 \pm (q-1)x]^{\pm \frac{1}{q-1}}\,, \quad \ln^{(\pm)}_q(x)=\pm \frac{x^{\pm(q-1)}-1}{q-1} \,,  \label{eq:eq_logq}
\end{equation}
where $^{(+)}$ in $e_q(x) \, (\ln_q(x))$ stands for $x \ge 0 \, (1)$, and~$^{(-)}$ stands for $x < 0 \, (1)$. The two versions of the functions in Equation~(\ref{eq:eq_logq}), denoted by $^{(+)}$ and $^{(-)}$, are in general needed when considering finite chemical potential. Notice that $e_q^{(+)}(x) = e_q(x)$ and $\ln_q^{(-)}(x) = \ln_q(x)$, where $e_q$ and $\ln_q$ were defined in Equations~(\ref{eq:logq}) and (\ref{eq:eq}), respectively. 
Then, the grand-canonical partition function for a non-extensive ideal quantum gas is given by~\cite{Megias:2015fra}
\begin{equation}
 \ln Z_q(V,T,\mu) =
 -\xi V\int \frac{d^3p}{{(2\pi)^3}} \sum_{r=\pm}\Theta(r x)\ln^{(-r)}_q \left( \frac{ e_q^{(r)}(x)-\xi}{ e_q^{(r)}(x)} \right)  \,, \label{eq:log_Zq}
\end{equation}
where $x = (E_p - \mu_Q)/(k_\B T)$, the particle energy is $E_p = \sqrt{p^2+m^2}$, with $m$ being the mass and $\mu_Q$ the chemical potential, $\xi = \pm 1$ for bosons and fermions, respectively, and~$\Theta$ is the step function. Note that $e_q^{(\pm)}(x) \stackrel[q \to 1]{\longrightarrow}{} e^x \quad $ and $\quad \ln_q^{(\pm)}(x) \stackrel[q \to 1]{\longrightarrow}{} \ln(x)$, so that Tsallis statistics reduces to B-G statistics in the limit $q \to 1$. As originally introduced in the literature, the $e_q^{\pm}(x)$ and $\ln^{(\pm)}_q(x)$ functions do not satisfy the Kubo-Martin-Schwinger (KMS) relations and show a break in the second derivatives at the Fermi surface (see Ref.~\cite{Biro:2014ata} and references therein). This issue was solved in Ref.~\cite{Megias:2015fra} by the introduction of some terms 
in the average number of particles,  $\langle N \rangle$ , and average energy, $\langle E \rangle$, that appear by taking the derivatives of the partition function of Eq.~(\ref{eq:log_Zq}) with respect to $\mu_Q$ and $T$, after performing the momentum integral. Later on, 
this gap was studied in detail in the search for physical significance~\cite{Rozynek:2016ykp,Rozynek:2019qzq}.

The thermodynamics of Tsallis statistics which follows from Equation~(\ref{eq:log_Zq}) has been used to successfully describe the thermodynamics of QCD in the confined phase by using the hadron resonance gas approach, with applications in high energy physics~\cite{Megias:2015fra}, hadron physics~\cite{Andrade:2019dgy}, and~neutron stars~\cite{Menezes:2014wqa}.

\section{Application and Results}
\label{sec:results}

%\subsection{Experimental evidences} \label{sect:experimentalevidences}

In this section, we will describe how the theory presented in Section~\ref{sec:Theoretical_Method} can describe many aspects of QCD as observed in different physical situations, in~particular in HEP. The~main result we obtained from the theory is an analytic formula for the effective coupling constant in the non-perturbative regime, which~follows as a consequence of the fractal structure. The~effects of fractality of the YMF extend to other aspects, as~the power-law behavior of the HEC distribution, which~are described by Tsallis statistics and allow to access experimentally the value of the entropic index, $q$, that results to be in very good agreement with the value found by the theory. We describe some aspects of the coupling constants and beta-function.

\subsection{Transverse Momentum Distributions}

\begin{figure}[h]%
\centering
 \begin{tabular}{c@{\hspace{3.5em}}c}
 \includegraphics[width=0.37\textwidth]{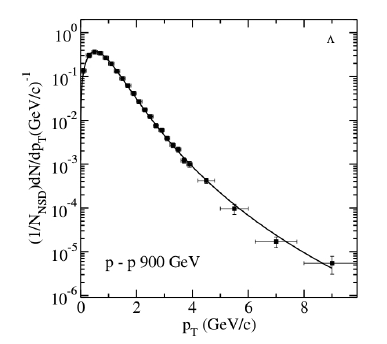} &
\includegraphics[width=0.31\textwidth]{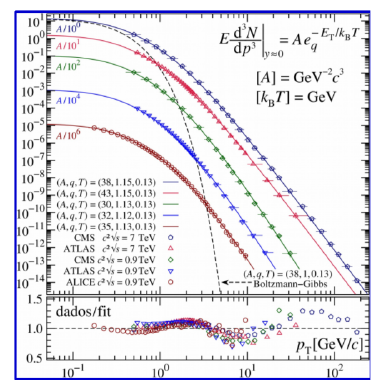} 
\end{tabular}
%\vspace{-4cm}
 \caption{Left panel: An example of fit of the calculated distribution to experimental data. Right panel: Fitting of the $p_\perp$ distribution formula to experimental data. Notice that the formula can fit well data over 15 orders of magnitude.}
\label{fig:figure_pT}
\end{figure}

The predictions of NESCT have been compared with HEP experimental data
for $p_\perp$ distributions (Equation~(\ref{eq:dNdpT})), showing a good
agreement between calculation and
data~\cite{Cleymans:2011in,Sena:2012ds,Marques:2012px,Azmi:2014dwa,Azmi:2015xqa},
resulting in $q = 1.14 \pm 0.01$ and $T = 62 \pm 5$~MeV. In addition, the hadron mass spectrum
formula has been used to describe the known hadronic states, resulting
again in a very good agreement with data and leading to values of $q$
and $T$ very similar to those obtained with $p_\perp$
analysis~\cite{Marques:2012px}. An example of fitting to $p_\perp$
distribution is shown in Figure~\ref{fig:figure_pT} (left).

Careful fittings of $p_\perp$ distribution power-law formulas to high
statistics data have shown some oscillations in the ratio between
calculated and experimental data, which~has already being attributed
to fractal aspects of the hadronic
interaction~\cite{Rybczynski:2012vj}, as~shown in
Figure~\ref{fig:figure_pT} (right). Notice also the wide range of experimental
data that can be fitted with power-law formula. This may indicate that
there are non-equilibrium events that are produced without the
formation of a thermodynamically equilibrated system, but the
multiplicity for such events must be smaller than 5. More
investigations in this direction would be welcome. 

\begin{figure}[t]
\centering
\includegraphics[width=0.46\textwidth]{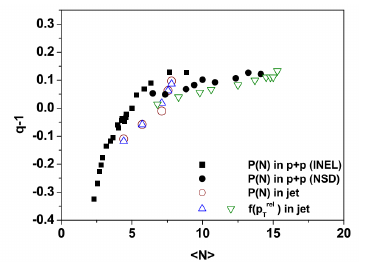}
 \caption{Behavior of the parameter $q$ as a function of
multiplicity. It~indicates the self-similarity between the system
formed at high energy collisions and the jets produced by that
system. In addition, it may indicate the region of multiplicity where
particle production without the formation of an equilibrated system
is relevant. Figure taken from Ref.~\cite{Wilk:2013jsa}.}
\label{fig:figure_qN}
\end{figure}

It is interesting the fact that the distributions for all particles
produced in collisions at different energies can be described by
roughly the same values for the parameters $T$ and $q$ of the non-extensive thermodynamics, as~shown in
Ref.~\cite{Marques:2012px}, cf. left panel of Figure~\ref{fig:experiment}. However, a study of the best fitted values
for those parameters as a function of the multiplicity has shown that
the parameters can be considered constant only for events with
multiplicity larger than 5 or 6, as~shown in the
Figure~\ref{fig:figure_qN}.

Another interesting aspect of the results shown in the Figure above is
that there is a clear self-similarity in the $p_\perp$ distributions of
particles and jets in relation to the beam direction, but there is
also the same distribution for particles forming a jet in relation to
the jet direction~\cite{Wilk:2013jsa}. This is maybe the most direct
experimental evidence of self-similarity in HEP and~should be studied
in more detail and with better statistics.

\subsection{Rapidity Distributions}

The use of the non-extensive thermodynamics allows a simple
description of most of the data from HEP. In~addition to the $p_\perp$
distribution described with just two parameters that have the same
value for any secondary particle and any collision energy, also the
rapidity distribution can be described by a simple model which
includes other two parameters to describe the movement of two
fireballs going apart after the collision. The~two parameters are the
position and width of two Gaussian functions describing the movement
of those two fireballs. The~two Gaussians are supposed to present some
symmetries, so they can be described by only two parameters. Such
model was presented in Ref.~\cite{Marques:2015mwa}, and the results
show that the parameters for those Gaussian are also approximately
constant for different collision energies for the range considered in
that work, as~shown in figures below. \mbox{In~the same work}, predictions
for particle production in $pp$ collisions at 13 GeV were made, but, up
to now, those predictions were not confronted with experimental data.

\subsection{Intermittency}

One of the most used techniques to unveil fractal dimensions from measured distributions is the intermittency method, that we briefly describe below.
Consider the collision of two systems, each of which may be a lepton,
a hadron, or a nucleus. Let ${\mathcal N}(y)$ be the rapidity ($y$) distribution of particles produced in an event at high energy. We
focus our attention on a narrow window of width~$\delta y$. The~averaging over many similar windows is a process that can be
considered later for the purpose of increasing the experimental
statistics but should be put aside for the present. Let $k$ denote the
number of particles detected in $\delta y$ in one event. Obviously, $k$ will
fluctuate from event to event. Let $Q_k$ be the number of times that
the multiplicity $k$ recurs in ${\mathcal N}$ events. Note that $Q_k$ is not
normalized, so it is not a distribution. Define~\cite{Bialas:1985jb,Bialas:1988wc} 
\begin{equation}
P_k = Q_k /{\mathcal N}\,,
\end{equation}
where 
\begin{equation}
{\mathcal N} = \sum_{k=0}^\infty Q_k  \,.
\end{equation}

The normalized moments are 
\begin{equation}
C_q = \frac{ \sum_{k=0}^\infty k^q P_k }{ \left( \sum_{k=0}^\infty k P_k  \right)^q } 
\end{equation}
where $q > 0$ is the momentum order. Another useful quantity is defined by~\cite{Hwa:1989vn}
\begin{equation}
G_q = \sum_{k=0}^\infty k^q Q_k/ K^q,
\end{equation}
where
\begin{equation}
K =  \sum_{k=0}^\infty k Q_k \,.
\end{equation}

The probability
\begin{equation}
P_i = k_i / K 
\end{equation}
represents the probability to have an event with $k_i$ particles in
the window with width~$\delta y$. For $\delta y$ sufficiently small, the
behavior of this probability with $\delta y$ is 
\begin{equation}
P_i \propto \delta y^\alpha  \,.
\end{equation}

\begin{figure}[h]
\centering
 \includegraphics[width=0.9\textwidth]{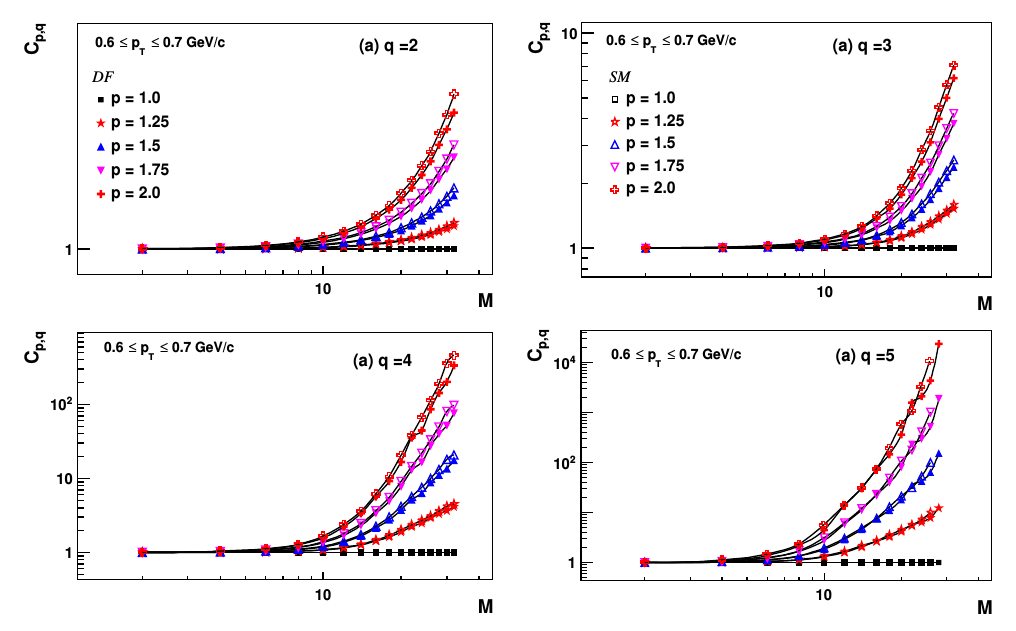}
 \caption{Analysis of moments in intermittency study for the $p_\perp$ window $0.6 \le p_\perp \le 0.7 \,\textrm{GeV}$, \mbox{cf. Ref.~\cite{Gupta:2015qja}}.}
\label{fig:figure_intermittency}
\end{figure}

\begin{figure}[h]
\centering
\vspace{-1.7cm}
\includegraphics[width=0.40\textwidth]{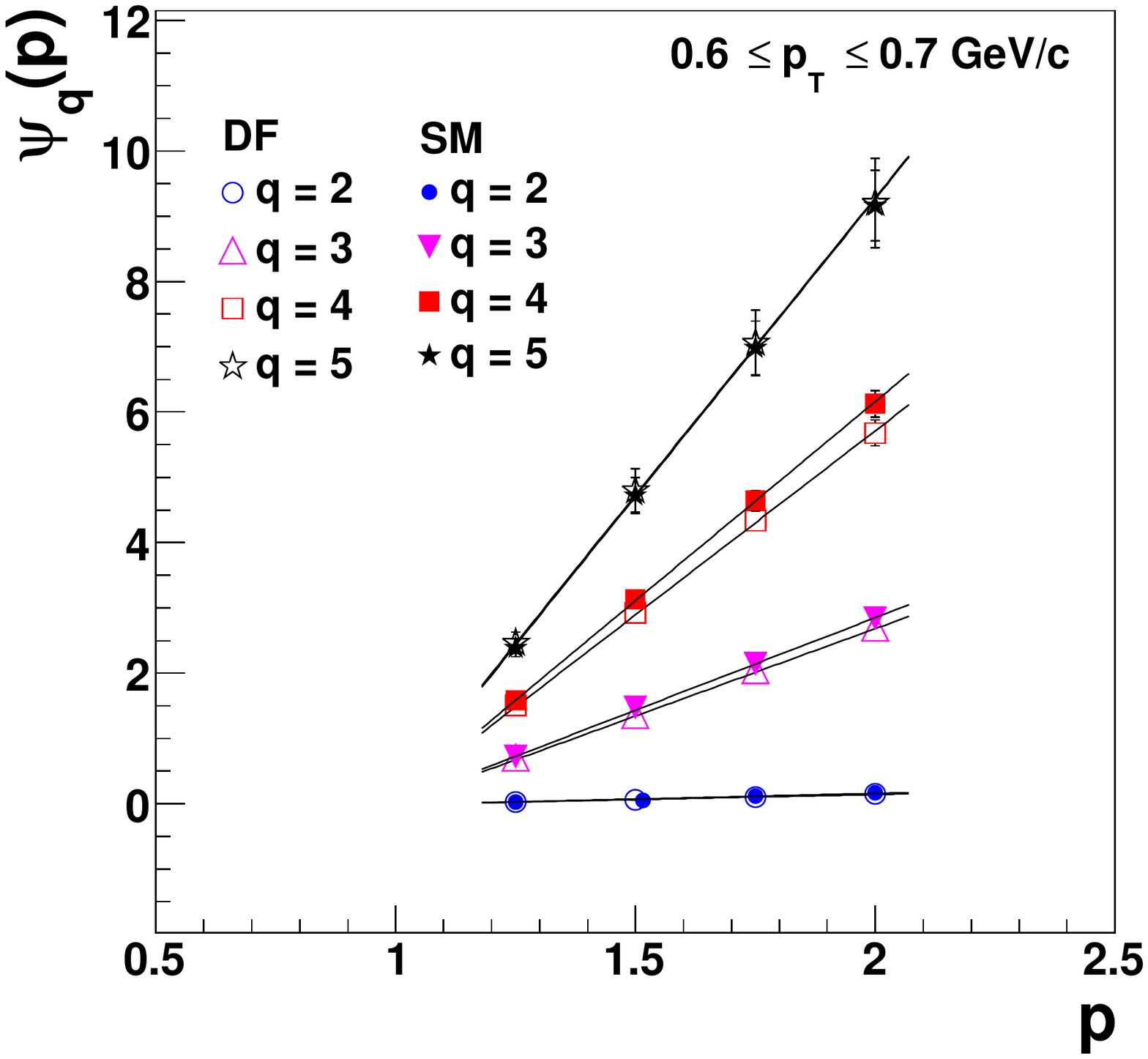}
\vspace{-1.5cm}
 \caption{$\psi_q$ defined in Equation~(\ref{eq:psi_q}) as a function of the momentum in intermittency study for the $p_\perp$ window $0.6 \le p_\perp \le 0.7 \,\textrm{GeV}$, cf. Ref.~\cite{Gupta:2015qja}.}
\label{fig:figure_psi_intermittency}
\end{figure}

The moments $C_q$ or $G_q$ depends on $\delta y$ as 
\begin{equation}
G_q \propto \delta y^{q\alpha - f(\alpha)}  \,,
\end{equation}
where $f(\alpha)$ is known as fractal spectrum. In addition,
\begin{equation}
G_q \propto \delta y^{\tau(q)} \,;
\end{equation}
therefore, $\tau(q) = q \alpha - f(\alpha)$. Note that we can make the identification
\begin{equation}
\tau(q) = (q - 1) D_q \,,
\end{equation}
where $D_q$ is the Hausdorff dimension at
order~$q$. The~advantage of the second momentum $(G_q)$ is that it is
not calculated for each event, but to a collection of events. A~nice
description of the experimental approach to the relevant quantities is
given in~\cite{Agababyan:1996vu}. A typical result of such analysis is
shown in Figures~\ref{fig:figure_intermittency} and
~\ref{fig:figure_psi_intermittency}. In~Figure~\ref{fig:figure_intermittency},
the cumulative moments (also called fractality moments) $C_{p,q} =
C_q(p)$ are displayed as functions of $\mathbb M$, where $\mathbb M = 1/\delta y$ is
the number of bins in which the range of momenta is considered in the
distribution, and~$\delta y$ is the bin width. The~$\mathbb M$-dependence of
$C_{p,q}$ seems to be close to a power-law behavior in $\mathbb M$, so that
one can also define the scaling exponent $\psi_q(p)$ as

\begin{equation}
C_{p,q}(\mathbb M) \propto {\mathbb M}^{\psi_q(p)} \,, \label{eq:psi_q}
\end{equation}
which is referred to as {\it erraticity}. In~Figure~\ref{fig:figure_psi_intermittency} it is shown that the dependence of $\psi_q$ with the momentum is linear for each value of $q$.

\unskip

From the thermofractal structure one can obtain the fractal dimension
of hadrons, what was done in Ref.~\cite{Deppman:2016fxs}, resulting in
$D = 0.69$, a value that is close to that resulting from intermittence
analysis, around $D = 0.65$~\cite{Azhinenko:1989jr,Albajar:1992hr,Sarkisian:1993mf,Rasool,Singh:1994uh,Ghosh:1998gw}. For a more complete description of intermittency and its application to HEP, see Refs.~\cite{Kittel,DeWolf:1995nyp}.

\subsection{Tsallis Statistics and QCD Thermodynamics}
\label{sec:thermodynamics_QCD}

The thermodynamics of QCD in the confined phase can be studied within
the HRG approach, which~is based on the assumption that physical
observables in this phase admit a representation in terms of hadronic
states which are treated as non-interacting and point-like
particles~\cite{Hagedorn:1984hz}. These states are taken as the
conventional hadrons listed in the review by the PDG~\cite{Zyla:2020zbs} .
 Within this
approach, the partition function is then given
by~\cite{Megias:2015fra,Menezes:2014wqa}

\begin{equation}
\ln Z_q(V,T,\{\mu_Q\})=\sum_i \ln Z_q(V,T,\mu_{Q_i})\,, \label{eq:logZ}
\end{equation}
where $Z_q(V,T,\mu_{Q_i})$ is the partition function of a
non-extensive ideal quantum gas given by Equation~(\ref{eq:log_Zq}),
$q$ is the entropic index, and~$\mu_i$ refers to the
chemical potential for the {\it i-th} hadron.
\begin{figure}[h]%[htb]
\centering
\includegraphics[width=6.7cm]{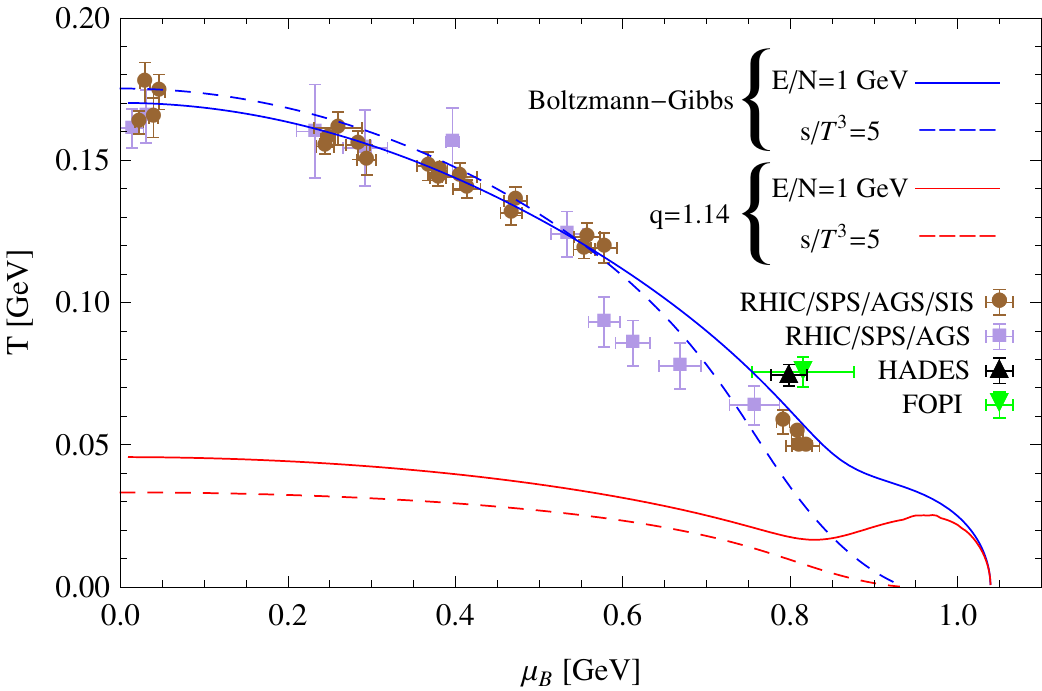} \hspace{1cm}
\includegraphics[width=6.7cm]{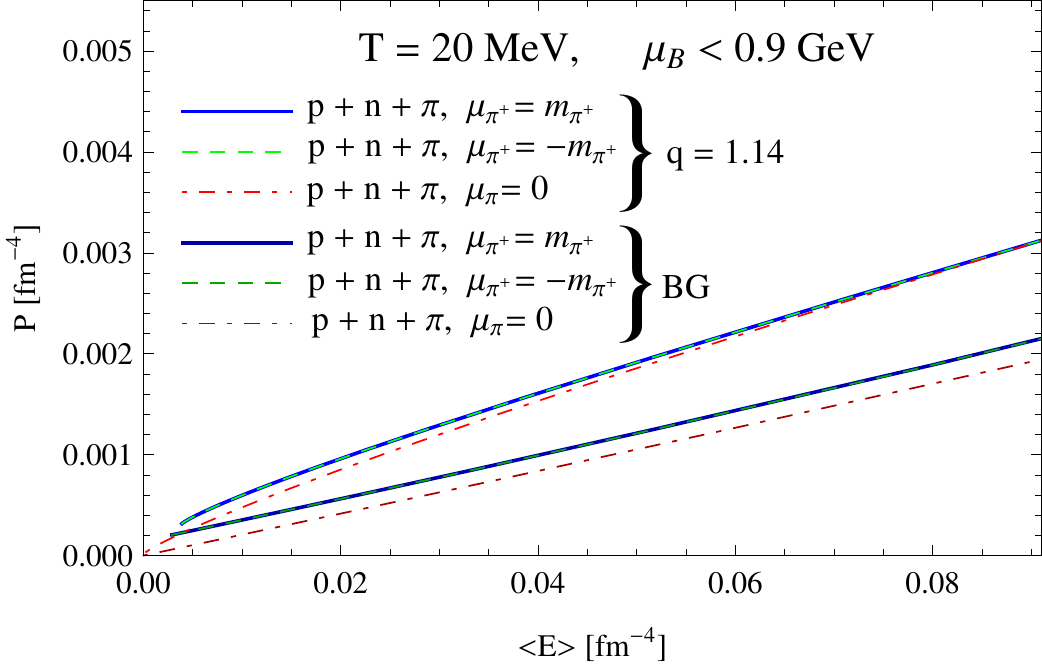} 
\caption{Left panel: Chemical~freeze-out~line~$T = T(\mu_B)$ obtained by assuming $\langle E\rangle / \langle N \rangle = 1 \GeV$ (continuous lines) and $s/T^3 = 5$ (dashed lines). We plot the result by using Boltzmann--Gibbs statistics, and~Tsallis statistics with $q=1.14$. Experimental data are taken from: Relativistic Heavy Ion Collider (RHIC)
/SPS
/Alternating Gradient Synchrotron (AGS)
/SIS
,~\cite{Cleymans:2005xv}, RHIC/SPS/AGS~\cite{Andronic:2005yp},HADES
~\cite{Agakishiev:2010rs}, and~FOPI
~\cite{Lopez:2007aa}. Right panel: EoS 
of hadronic matter. Figure taken from Ref.~\cite{Megias:2015fra}.} 
\label{fig:1}
\end{figure}

The thermodynamic functions can be obtained from Equation~(\ref{eq:logZ})
by using the standard thermodynamic relations. There are different
proposals for the conditions determining the transition line between
the confined and the deconfined regimes of QCD. Using the arguments
of~\cite{Cleymans:1999st}, the phase transition line in the $T - \mu_B$
diagram, with $\mu_B$ the baryonic chemical potential, can be determined by the condition $\langle E\rangle / \langle
N \rangle = 1 \GeV$, where the brakets stand for thermal expectation values. A~different method based on the entropy density
has been proposed in~\cite{Tawfik:2005qn}. The~result for the chemical
freeze-out line is displayed in Figure~\ref{fig:1} (left). The
transition line determines the region where the confined states exist
(below the line), and~the region where one expects to find the
quark-gluon plasma (above the line). We observe in this figure that
the freeze-out line spams over the region of $0<\mu_B<1039.2\,\MeV$
with a maximum value for $\mu_B$ corresponding to a null critical
temperature. The~curves show an inflection for $\mu_B \sim 0.9 \,
\GeV$ which is related to the sharp increase in the baryon density as
the baryonic chemical potential approaches the proton/neutron mass
$m_{p,n} \simeq 0.94 \, \GeV$. A~similar behavior is observed in
Boltzmann--Gibbs statistics. For $\mu_B=0$ the effective temperature is
$T_0 = 45.6\,\MeV$ for $q=1.14$, which~is not in full agreement with the
value $T_0 = 62 \pm 5 \, \MeV$ found in the analysis of the
$p_\perp$-distributions~\cite{Marques:2012px}. This disagreement can be
related to the value adopted for $\langle E \rangle/\langle N
\rangle$, which~still must be checked by analysis of experimental data
with the non-extensive statistic. In order to provide an estimate of
the sensitivity of the value of $T_0$ when changing $q$ it is worth
mentioning that the effective temperature for a slightly smaller value
of the entropic index, $q=1.12$, is $T_0 = 61.0\, \MeV$, which~is in
much better agreement with the $p_\perp$-distribution analysis. We also
display in Figure~\ref{fig:1} (right) the Equation of State (EoS) for
hadronic matter at finite baryonic chemical potential. It~it
remarkable that $P(E)$ becomes larger in Tsallis statistics as
compared to B-G statistics (see a discussion
in~\cite{Megias:2015fra,Menezes:2014wqa}). These thermodynamic studies
lead to the natural value $q = 1.14$, which~is in accordance with many
previous analysis, like the results from $pp$ collisions and the beta
function of QCD studied above. 

\begin{figure}[h]%[htb]
\centering
\includegraphics[width=7.7cm]{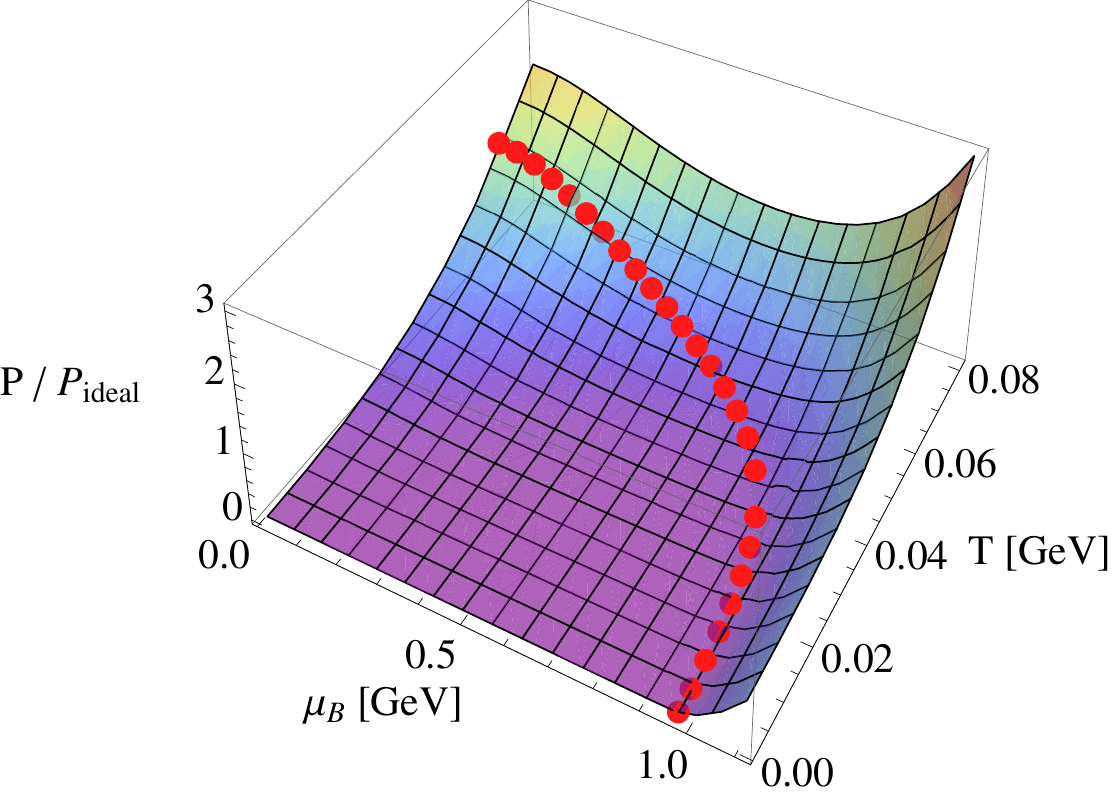} \hspace{1cm}
\includegraphics[width=5.7cm]{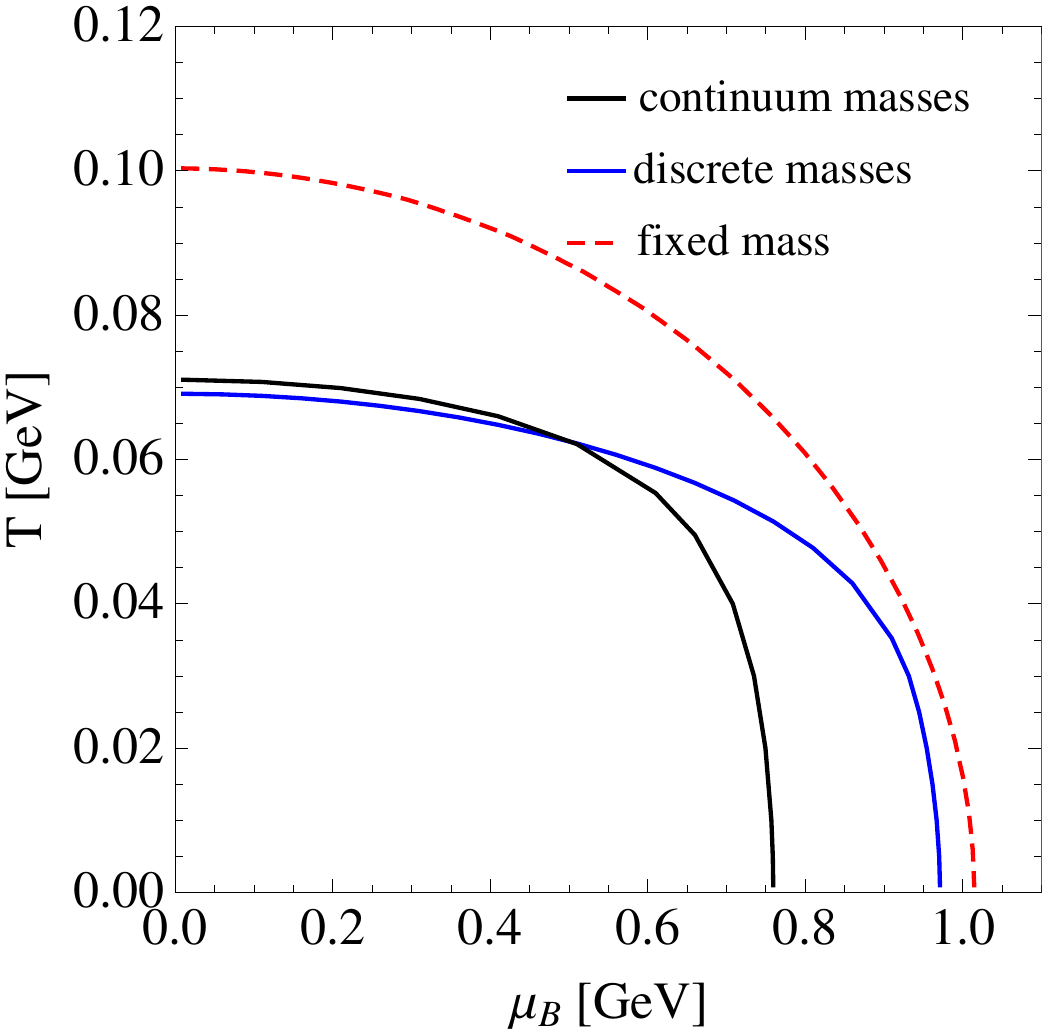} 
\caption{Left panel: Pressure (normalized to the ideal gas limit) as a function of temperature and baryonic chemical potential in the case of a non-extensive gas with particles with discrete masses. Red dots indicate the region where the gas total energy inside a volume $V_{\textrm{proton}}$ is equal the proton mass, i.e.,~$\varepsilon \cdot V_{\textrm{proton}} = m_{\textrm{proton}}$. Right panel: Temperature as a function of baryonic chemical potential~$T = T(\mu_B)$ for a non-extensive gas within the constraint $\varepsilon \cdot V_{\textrm{proton}} = m_{\textrm{proton}}$. For comparison, the results are displayed in three different scenarios: i) continuum masses, ii) discrete masses, and~iii) fixed mass. It~is considered the value $q = 1.14$ in both panels. Left panel taken from Ref.~\cite{Andrade:2019dgy}.}
\label{fig:TmuB_Bag}
\end{figure}

The QCD thermodynamics has been studied, as in Ref.~\cite{Andrade:2019dgy}, by using Tsallis statistics within the Massachusetts Institute of Technology (MIT) bag model. 
This allowed to investigate the underlying fractal structure of hadrons, leading to the emergence of non-extensivity of the hadronic thermodynamics. The~thermodynamic quantities have been studied in the approximation of fixed mass for all bag constituents, but also within the discrete mass approximation given by the PDG spectrum~\cite{Zyla:2020zbs}, 
as~well as the continuum spectrum provided by the NESCT through Equation~(\ref{eq:rho_m_2}). The~conclusion is that the continuum spectrum scenario fully applied the hypothesis that the hadron bag is an ideal gas of strongly interacting particles, and~this corresponds to the picture of the hadrons as thermofractals. It~is displayed in Figure~\ref{fig:TmuB_Bag} the regime in the phase space $(\mu_B,T)$ in which the proton exists as a bound particle, i.e.,~$\varepsilon \cdot V_{\textrm{proton}}  \leq m_{\textrm{proton}}$. The~results in the three scenarios mentioned above are displayed in the left panel of this figure. 

\subsection{Discussion} 
\label{sect:Applications}

The theoretical method developed above allows us to understand on solid grounds the emergence of Tsallis statistics in HEC. The~predictions of the theory find support on experimental data available, and~can explain, in~a single theoretical framework, many aspects of the HEP and of the hadronic systems. The~effective coupling obtained from considerations about scale symmetry of the YMF shows how four-momentum imparted to one effective parton is distributed among its internal d.o.f., which~are, themselves, associated to new effective partons created in the interactions between the fields.

The description of the results from high energy collisions by thermodynamics methods is interesting by itself because of possible applications in Astrophysics and in Cosmology. In~fact, a description of baryogenesis in the early universe must necessarily take into account the properties of the phase transition between confined and deconfined regimes of the hadronic matter.

The extension of the NESCT to systems with finite chemical
potential~\cite{Megias:2015fra} has allowed the application of the
thermodynamics properties of hadrons to the study of neutron stars
(NS) ~\cite{Lavagno:2011zm,Menezes:2014wqa}. While high energy
collisions take place at high temperatures and low chemical
potentials, protoneutron and NS are characterized by high chemical
potentials and low temperatures, in~a completely different region of
the QCD phase diagram. When applied to describe these compact objects,
the Tsallis statistics results on a very small change in the main
macroscopic properties, as~maximum mass and radius, but we found that
the internal temperature of the stars decreases with the increase of
the $q$-value (taken within a reasonable range) and, around the star
central densities the temperature decreases by approximately 25\% in
average. This aspect may have important consequences when the star
evolves from a hot and lepton rich object to a cold and deleptonized
compact star. Moreover, the direct Urca process is substantially
affected by non-extensivity, with probable consequences on the cooling
rates of the stars. Notice that some authors have considered $q < 1$
to study NS~\cite{Lavagno:2011zm}. However, in this case the stellar
maximum masses decrease with respect to the B-G statistic analysis, so
that in subsequent studies the case $q > 1$ only was taken into
account. Besides NS constituted of hadrons, strange stars constituted
of deconfined quarks were analyzed with a simple model
~\cite{Cardoso:2017pbu}. The~conclusions discussed above were all
corroborated and the construction of the QCD phase diagram first order
transition has shown that the limiting curve is quite sensitive to non-extensivity. An investigation with a model that allows the calculation
of the critical end point (CEP) remains to be done.

The interpretation of the power-law
distributions by the thermofractal structure offers the possibility
to connect such structure to QCD properties, as~shown in
Ref.~\cite{Deppman:2016fxs}. Finally using the blast-wave approach
to the expansion of the quark-gluon plasma, one can extend the present
thermodynamics theory so as to apply to the problem of high energy
nucleus-nucleus collisions, as well, for which there is a large number
of accurate data~\cite{Feal:2018ptp}.

Finally, the hadron structure can benefit from the theory presented here. In~fact, some models have already used Tsallis statistics as a way to introduce the fractal aspects of the QCD in the description of hadron structure~\cite{Andrade:2019dgy}.

\section{Conclusions}
\label{sect:Conclusions}

In this work, we reviewed the applications of Tsallis statistics to HEP, hadron physics, and astrophysics. We have investigated the structure of a thermodynamical system presenting fractal properties showing that it naturally leads to Tsallis non-extensive statistics. Based on the scaling properties of gauge theories and thermofractal considerations, we have shown that renormalizable field theories lead to fractal structures, and~these can be studied with Tsallis statistics. By using a recurrence formula that reflects the self-similar features of the fractal, we have computed the effective coupling and the corresponding beta function. The~result turns out to be in excellent agreement with the one-loop beta function of QCD. Moreover, the entropic index, $q$, has been determined completely in terms of the fundamental parameters of the field theory. Finally, the result for $q$ is shown to be in good agreement with the value obtained by fitting Tsallis distributions to experimental data. \mbox{These results give} a solid basis from QCD to the use of non-extensive thermodynamics to study properties of strongly interacting systems and, in~particular, to use thermofractal considerations to describe hadrons. In~the last part of the manuscript we have discussed the experimental results on transverse momentum and rapidity distributions in $pp$ collisions that provide a solid basis to study the role of Tsallis in HEP, as~well as a discussion of some applications including the QCD thermodynamics, the QCD phase diagram and the physics of proto(neutron) stars.

From what was discussed here, one can see that there are still many
open questions regarding the description of HEP data by power-law
distributions. The~main problem is to verify to what extent the idea
of fractal structure can describe experimental data. In~this regard,
not only the $p_\perp$ distributions must be carefully analyzed, but also
other quantities, mainly the fractal dimension that can be accessed
through intermittency analysis. Although this analyses are still
performed for HEP collisions on emulsion, no such analysis was
performed with LHC data. Aside intermittency, the study of the
parameters $T$ and $q$ as a function of multiplicity can shed some light
on the discussion about the nature of the power-law distributions: is
it due to a thermodynamical system in equilibrium or is it motivated
by QCD aspects? If it is shown with accuracy that for multiplicities
larger than 5 or 6 the values are constant, as~predicted by the
thermodynamical theory, this would favor the thermodynamics
interpretation. The~same analysis would allow for a discussion on
the self-similarity in experimental data, as~already discussed above.

One should keep in mind that the value for the entropic index, $q=1.14$, obtained from the QCD parameters is a constant related to the fundamental parameters of the QCD, namely the number of colors and number of flavors. However, the assumption that the configuration space for any effective parton is completely available holds only under some special situations, as~those found in HEP collisions. Whenever further constraints in the possible color and flavor, or even topological restrictions of the momentum space, are present, the number of d.o.f. will vary and therefore also the value for $q$ observed in those situations. The~effects of constraints are more evident when the particle multiplicity is small, or when experimental setup is such that just specific particle configurations are selected. Such cases can happen at low energy collisions and in very peripheral collisions.

The rapidity distribution can be further analyzed, and~the predictions
made for $pp$ collisions at 13 GeV can be compared with the data
already measured at that energy. In~addition, the theory can be
extended to account for nucleus-nucleus collision, with the inclusion
of the blast-wave model in the thermodynamical description, allowing
for the analysis of $AA$ data. In summary, the interesting analysis
that could be made with recent experimental data are (initially for
$pp$ but eventually \mbox{also for $AA$):}
\begin{enumerate}[leftmargin=*,labelsep=4.9mm]
\item Analysis of $p_\perp$ distribution with high
statistics, obtaining the behavior of $T$ and $q$ as a function of
multiplicity for different particles. 
\item Investigation of self-similarity in jet and secondary particle
distributions.
\item Investigation of the rapidity distribution as predicted with the
theory, and~also of the distributions for 13 GeV.
\item Investigation of intermittency in HEP data from LHC.
\end{enumerate}

These and other issues will be addressed in a forthcoming publication.

%%%%%%%%%%%%%%%%%%%%%%%%%%%%%%%%%%%%%%%%%%
\vspace{6pt} 

%%%%%%%%%%%%%%%%%%%%%%%%%%%%%%%%%%%%%%%%%%
%% optional
%\supplementary{The following are available online at \linksupplementary{s1}, Figure S1: title, Table S1: title, Video S1: title.}

% \supplementary{The following are available at \linksupplementary{s1}, Figure S1: title, Table S1: title, Video S1: title. A~supporting video article is available at doi: link.}

%%%%%%%%%%%%%%%%%%%%%%%%%%%%%%%%%%%%%%%%%%
\authorcontributions{Conceptualization, A.D.; methodology, A.D., E.M. and D.P.M.; software, A.D. and E.M.; validation, A.D. and E.M.; formal analysis, A.D. and E.M.; investigation, A.D., E.M. and D.P.M.; resources, A.D. and E.M.; data curation, A.D. and E.M.; writing---original draft preparation, A.D., E.M. and D.P.M.; writing---review and editing, A.D., E.M. and D.P.M.; visualization, A.D. and E.M.; supervision, A.D.; project administration, A.D. and E.M.; funding acquisition, A.D., E.M. and D.P.M. All authors have read and agreed to the published version of the manuscript.} 

%%%%%%%%%%%%%%%%%%%%%%%%%%%%%%%%%%%%%%%%%%
\funding{The research of A.D. and D.P.M. were funded by the Conselho Nacional de
  Desenvolvimento Cient\'{\i}fico e Tecnol\'ogico (CNPq-Brazil) and by
  Project INCT-FNA Proc. No. 464898/2014-5, and~by FAPESP under grant
  2016/17612-7 (A.D.). The~work of E.M. is funded by the Spanish MINEICO
  under Grant FIS2017-85053-C2-1-P, by the FEDER Andaluc\'{\i}a
  2014-2020 Operational Programme under Grant A-FQM-178-UGR18, by
  Junta de Andaluc\'{\i}a under Grant FQM-225, by the Consejer\'{\i}a
  de Conocimiento, Investigaci\'on y Universidad of the Junta de
  Andaluc\'{\i}a and European Regional Development Fund (ERDF) under
  Grant SOMM17/6105/UGR, and~by the Spanish Consolider Ingenio 2010
  Programme CPAN under Grant CSD2007-00042. The~research of E.M. is
  also supported by the Ram\'on y Cajal Program of the Spanish MINEICO
  under Grant RYC-2016-20678.} 

%%%%%%%%%%%%%%%%%%%%%%%%%%%%%%%%%%%%%%%%%%
\acknowledgments{We would like to thank Constantino Tsallis for suggesting us to write this review.} 

%%%%%%%%%%%%%%%%%%%%%%%%%%%%%%%%%%%%%%%%%%
\conflictsofinterest{The authors declare no conflict of interest.} 

%%%%%%%%%%%%%%%%%%%%%%%%%%%%%%%%%%%%%%%%%%

%%%%%%%%%%%%%%%%%%%%%%%%%%%%%%%%%%%%%%%%%%
%% optional
\appendixtitles{no} % Leave argument "no" if all appendix headings stay EMPTY (then no dot is printed after "Appendix A"). If the appendix sections contain a heading then change the argument to "yes".
%\appendix
%\section{}
%\unskip
%\subsection{}
%The appendix is an optional section that can contain details and data supplemental to the main text. For example, explanations of experimental details that would disrupt the flow of the main text, but nonetheless remain crucial to understanding and reproducing the research shown; figures of replicates for experiments of which representative data is shown in the main text can be added here if brief, or as Supplementary data. Mathematical proofs of results not central to the paper can be added as an appendix.
%
%\section{}
%All appendix sections must be cited in the main text. In~the appendixes, Figures, Tables, etc. should be labeled starting with `A', e.g., Figure A1, Figure A2, etc. 

%%%%%%%%%%%%%%%%%%%%%%%%%%%%%%%%%%%%%%%%%%
\reftitle{References}

% Please provide either the correct journal abbreviation (e.g.,~according to the “List of Title Word Abbreviations” http://www.issn.org/services/online-services/access-to-the-ltwa/) or the full name of the journal.

%\externalbibliography{yes}
%\bibliography{refs.bib}

%=====================================
% References, variant B: internal bibliography
%=====================================
%\begin{thebibliography}{999}

%\end{thebibliography}

% The following MDPI journals use author-date citation: Arts, Econometrics, Economies, Genealogy, Humanities, IJFS, JRFM, Laws, Religions, Risks, Social Sciences. For those journals, please follow the formatting guidelines on http://www.mdpi.com/authors/references
% To cite two works by the same author: \citeauthor{ref-journal-1a} (\citeyear{ref-journal-1a}, \citeyear{ref-journal-1b}). This produces: Whittaker (1967, 1975)
% To cite two works by the same author with specific pages: \citeauthor{ref-journal-3a} (\citeyear{ref-journal-3a}, p. 328; \citeyear{ref-journal-3b}, p.475). This produces: Wong (1999, p. 328; 2000, p. 475)

%%%%%%%%%%%%%%%%%%%%%%%%%%%%%%%%%%%%%%%%%%
%% optional

%% for journal Sci
%\reviewreports{\\
%Reviewer 1 comments and authors’ response\\
%Reviewer 2 comments and authors’ response\\
%Reviewer 3 comments and authors’ response
%}

%%%%%%%%%%%%%%%%%%%%%%%%%%%%%%%%%%%%%%%%%%
\end{document}